\documentclass[smallextended]{svjour3}       

\RequirePackage{fix-cm}

\usepackage{amsfonts}
\smartqed 
\usepackage{graphicx}
\usepackage{amssymb}
\usepackage{tablefootnote}
\usepackage{multirow}
\usepackage{censor}
\usepackage{amsmath}
\usepackage{apacite}
\usepackage{natbib}
\newcommand{\tabref}[1]{Table~\ref{#1}}
\newcommand{\figref}[1]{Figure~\ref{#1}}

\RequirePackage{color}
\definecolor{RED}{rgb}{1,0,0}
\definecolor{BLUE}{rgb}{0,0,1}
\definecolor{White}{rgb}{1,1,1}

\newcommand{\hide}[1]{}
\usepackage{setspace}

\setcitestyle{aysep={}} 
\usepackage[norule,hang]{footmisc}
\begin{document}

\title{Quantifying the relationship between student enrollment patterns and student performance
}


\author{Shahab Boumi         \and
        Adan Vela         
}


\institute{S. Boumi \at
              4000 Central Florida Blvd, Orlando, Florida \\
              Tel.: +1407-684-6694\\
              \email{sh.boumi@knights.ucf.edu}           
           \and
           A. Vela \at
              4000 Central Florida Blvd, Orlando, Florida\\
              \email{adan.vela@ucf.edu}
}

\date{Received: date / Accepted: date}
\titlerunning{Impacts of students' enrollment pattern on academic performance}
\maketitle
\begin{abstract}
Simplified categorizations have often led to college students being labeled as full-time or part-time students.  However, at many universities student enrollment patterns can be much more complicated, as it is not uncommon for students to alternate between full-time and part-time enrollment each semester based on finances, scheduling, or family needs.  While prior research has established full-time students maintain better outcomes then their part-time counterparts, limited study has examined the impact of enrollment patterns or strategies on academic outcomes.  In this paper, we applying a Hidden Markov Model to identify and cluster students' enrollment strategies into three different categorizes: full-time, part-time, and mixed-enrollment strategies. Based the enrollment strategies we investigate and compare the academic performance outcomes of each group, taking into account differences between first-time-in-college students and transfer students. Analysis of data collected from the University of Central Florida from 2008 to 2017 indicates that first-time-in-college students that apply a mixed enrollment strategy are closer in performance to full-time students, as compared to part-time students.  More importantly, during their part-time semesters, mixed-enrollment students significantly outperform part-time students.  Similarly, analysis of transfer students shows that a mixed-enrollment strategy is correlated a similar graduation rates as the full-time enrollment strategy, and more than double the graduation rate associated with part-time enrollment.  Such a finding suggests that increased engagement through the occasional full-time enrollment leads to better overall outcomes.

\end{abstract}

\section{Introduction}
%
%
%
%
%
%
%
%
%
%
%

In practice, either through choice or necessity, students engage in a variety of enrollment patterns over their academic career that includes full-time and part-time enrollment, or halting \citep{otoole_LongitudinalAnalysisFrequency_2003,hearn_EmergingVariationsPostsecondary_1992,cabrera_PathwaysFouryearDegree_2003,reardon_RaceIncomeEnrollment_2012,hearn_AttendanceHighercostColleges_1988}.
Based on a survey conducted at 253 academic institutions, only 29\% of students maintain full-time status over every semester they are enrolled, while 18\% of students maintain part-time enrollment over their entire academic career.  Meanwhile, most students, 53\%, alternate their enrollment status between part-time and full-time at least once during their studies \citep{center2017even}.  

To date, part-time enrollment status has been indicated as a risk-factor to student success. \cite{feldman1993factors} shows that, on average, full-time college students have higher retention rates and GPAs at the end of the first academic year than part-time students. In another study, \cite{pelkey2011factors} analyzed how race, age, enrollment status, GPA, and financial aid can impact students' persistence. Their analysis indicated that GPA and enrollment status have the highest impact on persistence at college.  Not only is enrollment status a factor, but so is course-load; as demonstrated in a study by \cite{darolia2014working}, students with more credits during their first semester are more likely to complete all the required credits for their program and ultimately their degrees.




Although enrollment status is perceived as critical to student success, there is no clear definition of what it means to be a \textit{part-time} student or \textit{full-time} student outside the ephemeral academic designation given each semester.  As most students intersperse full-time and part-time enrollment statuses, such a binary designation appears to be overly simplistic means to group students together for analysis based on their enrollment status during a single semester; we believe there is value in understanding more complex enrollment patterns and their relationship to student outcomes.  This assertion is supported by a 2015 nation-wide study indicating that student success can be found through mixed enrollment strategies \citep{track2015national} -- the authors report that non-first-time-in-college students that attend college utilizing a combination of part-time and full-time enrollment are less likely to drop out and more likely to complete degrees when compared to full-time students.     

In this study, we seek to find a more comprehensive means of identifying and grouping students according to their enrollment strategy (e.g., part-time, full-time, etc.).  Unlike a single-period model in which the students' strategy is equivalent to the observed semester enrollment status (i.e., part-time or full-time), we use a multi-period dynamic approach using a Hidden Markov Model (HMM). Through application of the HMM model a major contribution of this paper is a richer understanding of student enrollment patterns. Our model extends on traditional categorizations to include not only full-time and part-time enrollment strategies but also a mixed enrollment strategy.  Students who use a mixed enrollment strategy regularly intersperse full-time and part-time semesters.  After categorizing students into one three groups (i.e. full-time, part-time, and mixed enrollment strategy) we examine student academic outcomes (e.g., GPA, and DFW and graduation rates) associated with each strategy.  Accordingly, our second research contribution is a better understanding of the correlations between academic engagements (via enrollment strategies) and different measures of academic success.



\section{Problem statement}
We consider the problem of classifying students according to a higher-order  enrollment strategy as opposed to their enrollment status during any given semester.  For many students, the distinction between enrollment strategy and actual enrollment is minor. Students are considered full time in a given semester at the University of Central Florida if they enroll in more than 12 credits in that semester. Approximately 35\% of the student body at the University of Central Florida consistently enrolls full-time throughout their academic career, implying they employ a \textbf{strategy} of enrolling full-time. In contrast, the case for so-called \textit{part-time} students it is not so clear.  In any given semester, about 30\% of enrollments are part-time, yet only 7\% of students consistently enroll part-time over their academic careers. Enrolling part-time in a single semester over an academic career is not equivalent to the strategy of consistently enrolling part-time. It follows that a student who enrolls in a single semester part-time may not share similarities with other students who consistently enroll part-time.

This paper aims to recognize and report the distinction between a student's semester-by-semester enrollment status and enrollment strategy and find a more meaningful way to classify students' enrollments over their academic career.  More specifically, this paper develops a model that takes its input a sequence of enrollment statuses and returns a sequence of estimated strategies applied over the same time-frame.  In recognition that students use a greater diversity of strategy than just a full-time enrollment strategy (FES) or part-time enrollment strategy (PES), we introduce the notion of a mixed enrollment strategy (MES). For a mixed enrollment strategy, students may occasionally alternate between part-time and full-time enrollments. 

\tabref{tab:sampleEnrollmentData} provides examples of the enrollment status of four different students over their academic career along with the corresponding enrollment strategies. For instance, enrollment strategies for student number 1 through number 3 are FES, MES, and PES, respectively. Student number 1 has registered as a full-time student for five semesters and just one semester as part-time. Based on the model proposed in this paper, this student is classified as an FES student. From the example, one part-time semester among many full-time semesters does not affect a student's overall enrollment strategy.
On the other hand, student number 2 frequently alternates his enrollment status between full-time and part-time enrollment semester by semester. Therefore, the corresponding enrollment strategy for this student is MES.  Student number 4 applies two different enrollment strategies over their academic career. This student is classified as employing an ``other'' enrollment strategy. 

\begin{table}
\caption{Example enrollment status' over academic career and corresponding enrollment strategies}
\label{tab:sampleEnrollmentData}       
\centering
\begin{tabular}{lll}
\hline\noalign{\smallskip}
Student Number & Enrollment Status & Enrollment Strategy  \\
\noalign{\smallskip}\hline\noalign{\smallskip}
~~~~~~~~~~1 & F,P,F,F,F,F & F,F,F,F,F,F\\
~~~~~~~~~~2 & F,P,F,P,F,P & M,M,M,M,M,M\\
~~~~~~~~~~3 & P,F,P,P,F,P,P & P,P,P,P,P,P\\
~~~~~~~~~~4 & P,F,F,P,P,P,P & M,M,M,P,P,P,P \\
\hline 
 Legend & FT=F,~PT=P & FES=F,~MES=M,~PES=P\\
\noalign{\smallskip}\hline
\end{tabular}
\end{table}

\section{Literature review}
In practice, not every college registration leads to degree completion. Many students leave college before graduation with no degrees earned, imposing their families at a high cost \citep{garibaldi2012college,ryan2016educational}. There are verity factors that can affect degree completion rate at universities and cause delayed college enrollment and impact the time to graduation and part-time enrollment rate \citep{fagioli2015changing}. This section reviews some of the previous research investigating the effect of such factors, including student demographic characteristics, students' educational behavior patterns, and student engagement on their academic performance. We divide the literature into three main categories: students' socioeconomic backgrounds, students' academic behavior patterns, and students' engagement.

Many studies have shown that students' socioeconomic backgrounds can have a significant impact on student academic performance. \cite{goldrick2011accounting} claim that students who come from families with low-level income are more likely to leave college without degrees. The study also shows that offering additional financial aid to these students can significantly help them finish their program over the regular course of four years. Similarly, some other studies show that students with low socioeconomic backgrounds are more likely to leave university or have a longer time gap for a transition from high school to university \citep{cox2016complicating,goldrick2016reducing,rowan2007predictors,wells2012delayed}. The other demographic features that have been perceived as important factors affecting students' performance include race \citep{desjardins2006effects}, age \citep{jacobs2002age}, life course transition including work and marriage \citep{roksa2012late}, and gender \citep{taniguchi2005degree}. These studies showed that these factors could create inequality among college students in terms of academic success \citep{milesi2010all}.

Some other researchers investigated the relationship between students’ academic behavior patterns/trajectories and academic success \citep{boumi2019application,milesi2010all}. \cite{attewell2012academic} used the \emph{academic momentum perspective} concept to study the effect of students’ academic trajectories on the chance of degree completion. Their results show that the number of credits that students earn in the first semester creates a trajectory that has a positive correlation with degree completion probability. In another study, \cite{adelman2005moving} divided students' attendance pattern into three categories: \emph{Visitors}, \emph{Tenants}, and \emph{Homeowners}. \emph{Visitors} students acquire a few credits at community colleges and constitute the largest group through this classification. \emph{Tenant} students earn considerable credits, but in different colleges. Finally, the \emph{Homeowners} group consists of students with significant earned credits in community colleges. Similar classifications in students’ attendance patterns have been conducted in other research  \citep{marti2008dimensions,bahr2010bird,crosta2014intensity}. For example, the Adelman's \emph{Visitors} group is classified as \emph{one term and out} and \emph{early leavers} in the \cite{marti2008dimensions} and \cite{crosta2014intensity}, respectively, which similarly constitute the largest group of students. In these research, also a variety of attendance patterns is investigated. For example, 35\% of students in Crosta’s classification have been grouped as \emph{early leavers} students, who enroll for two or three semesters and then leave the college. In the research, the educational outcomes of students with these different attendance patterns are investigated. The study shows a positive correlation between students’ enrollment persistence and a community college credential. For instance, only 1\% of the \emph{early leavers} group students earn a credential within 5 or 6 years.

Over the past decade, student engagement has gained significant attention among education researchers. Many studies have indicated that there is a strong relationship between student academic performance and student engagement. \cite{kuh2009student} defines engagement as the amount of time students spend in collaboration with educational facilitates to enhance their knowledge and learning. The \emph{Community College Survey of Student Engagement} (CCSSE) defines five student engagement benchmark \citep{mcclenney2012student}: active and collaborative learning, student effort, academic challenge, student-faculty interaction, and support for learners. Some studies are focused on investigating the relationship between students' engagement and students' outcomes \citep{lee2014relationship,boumi2019application}. Most of the research conducted with CCSSE data demonstrate that the higher is student engagement, the better is the students' academic performance, including average GPA, retention and graduation rate, and persistence. For example, \cite{price2014student} provides empirical evidence that shows student engagement of type \emph{active and collaborative learning} has a significant impact on student graduation rate.

All these findings collectively show that a wide range of factors can create inequality among college students in their pathway toward academic success. A large number of these studies focus on the impact of students' engagement and educational behavior patterns. However, all such papers consider students' enrollment as their realized registration status (full-time or part-time), ignoring their inherent long-term enrollment behavior. In this paper, we model students' enrollment behavior by investigating the sequence of their enrollment status. By doing so, we classify students in three main categories: full-time enrollment strategy (FES), mix enrollment strategy (MES), and part-time enrollment strategy (PES). This new framework can tackle the previous research gap, which offers no clear definition for labeling a student as full-time or part-time through his educational career. Since the student enrollment strategy is not directly observable, we take advantage of Hidden Markov Modeling tools to tackle the proposed framework's inherent complexities.

\section{Methodology}
In this study, we generate and apply a Hidden Markov Model (HMM) to identify students' enrollment strategy and characterize the impact of enrollment strategy on student outcomes.  The use of HMM is not new to educational data mining and modeling. Previously it has been used to investigate students' sequential behaviors, decision-making, and performance  \citep{abdipredicting,bealmodeling,boyer2011investigating,halpern2018knowledge,hoernle2018modeling,falakmasir2016data}. As an example, \cite{balakrishnan2013predicting} classified students engagement (in terms of checking their course progress) in a massive open online course using Hidden Markov models and then predicted students retention for each class. Results show that students who never or scarcely check their progress are more likely to drop the course than students who actively check their progress. Other papers have used HMMs to model sequential student behavior. \cite{boyer2011investigating} modeled high school students' actions and behaviors using HMMs. By estimating HMM parameters with the Baum-Welch algorithm for each student, the authors clustered the students based on the individual transition matrices to assess differences in behavior and achievement between clusters. This study aimed to investigate the relationship between students' educational outcomes and learning tutorial modes. The hidden states in the research are defined as tutoring modes. The results of this study identify a significant correlation between tutorial dialogue structure and student learning outcomes. 

\begin{figure}
\centering
  \includegraphics[width=0.75\textwidth]{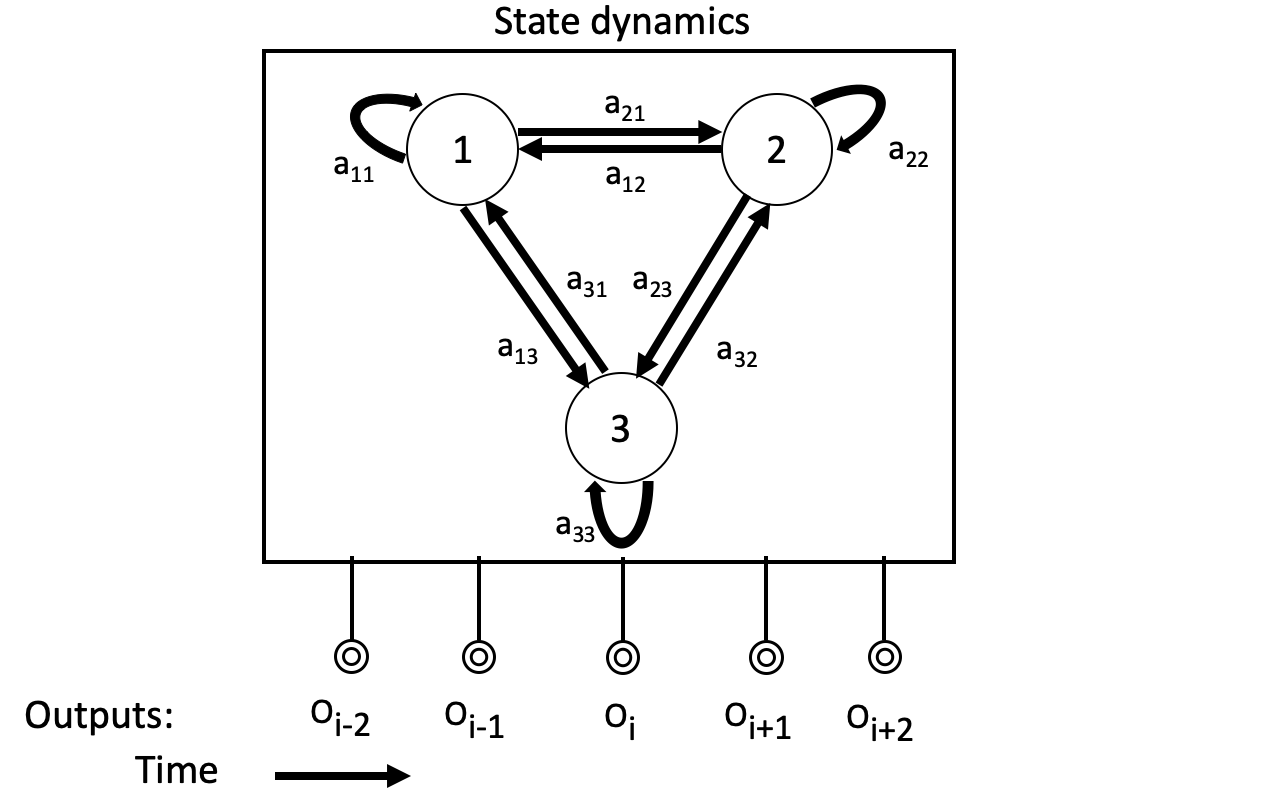}
\caption{Representation of a simple Hidden Markov Model}
\label{fig:HMM}       
\end{figure}

As depicted in \figref{fig:HMM}, similar to ordinary Markov Models, an HMM represents the dynamics of a system as it moves between operating states or modes (e.g., Modes 1, 2, and 3 in the figure). When operating within a state or mode, the system generates state-related output $O_i$  at each time-step. Unlike Markov Models, in the case of the standard HMM problem, the states are not always directly observable, and as such, they can only be inferred by observing a sequence of outputs. For the problem under consideration here, the hidden state corresponds to the enrollment strategy of a student (i.e., full-time enrollment strategy, mixed enrollment strategy, and part-time enrollment strategy), while the observations refer to the actualized enrollment in any given semester as reported in the student's academic history.


To give a formal definition of Hidden Markov models, we begin with the following notations: $Q=\{q_1,q_2,\hdots,q_N\}$ represents the set of $N$ possible states in the system; $A=[a_{i,j}]\in\mathcal{R}^{NxN}$ is a transition matrix, where each $a_{ij}$ denotes the probability of transitioning from state $i$ to $j$ at any given time-step; $O=o_1,o_2,\hdots,o_T$ represents a sequence of observations of length $T$, each drawn from the set of $M$ possible observations $V=\{v_1,v_2,\hdots,v_M\}$; and $\pi_o$ represents the initial distribution of the system. When a system is operating in a specific state $q_i$, the output $o_t$ at time $t$ is generated according to a unique probability distribution denoted as $B=b_i(o_t)$, the emission probability.

To generate an HMM to represent student enrollment strategies, we must learn the optimal model parameters $\lambda^*=(A,B,\pi_0)$ that reproduce known observations.  The process of learning $\lambda^*$ is based on the Baum-Welch algorithm, an iterative process that requires calculating the likelihood of any sequence of observations given $\lambda^*$, and decoding relationships between observations and hidden variables.  As the model is iteratively updated, the likelihood calculations and the decoding is updated.

\section{Student data records} 
The study presented in this paper makes use of processed undergraduate student records collected from the University of Central Florida, a large public university in the southeast United States, between the years 2008 to 2017. The total data-set amounts to approximately $170,000$ records. The data set contains a wide variety of information about students at UCF, including but not limited to: (1) demographic information, (2) admission information for students who have been admitted and enrolled, (3) degrees awarded (for bachelor level), (4) courses taken by students at UCF, and (5) FAFSA reported family income. Some of the self-reported demographic information along with the percentage of students who enroll as full-time and part-time, and admission type (FTIC and transfer) are provided in \tabref{tab:Gender} through \tabref{tab:Full-time part time semesters}. Unique to UCF, the student body includes a significant population of transfer students and Hispanic students\tablefootnote{UCF has recently been designated to be a Hispanic serving institution by the US Department of Education.}.
\begin{table}
\caption{Students gender distribution at UCF over 10 years}
\centering
\label{tab:Gender}       
\begin{tabular}{lll}
\hline\noalign{\smallskip}
{}&Females&Males\\
\noalign{\smallskip}\hline\noalign{\smallskip}
Percentage&56.2\% & 43.8\% \\
\noalign{\smallskip}\hline
\end{tabular}
\end{table}

\begin{table}
\caption{Students ethnicity distribution at UCF over 10 years}
\label{tab:Ethnicity}       
\centering
\begin{tabular}{lllll}
\hline\noalign{\smallskip}
{}&White&Hispanic&African-Am.&Other\tablefootnote{The other category includes American-Indian, Asian, Native Hawaiian, and Multi-racial ethnicity.}\\
\noalign{\smallskip}\hline\noalign{\smallskip}
Percentage&~55.2\% & ~~23.4\%&~~~~11.3\%&~10.1\% \\
\noalign{\smallskip}\hline
\end{tabular}
\end{table}

\begin{table}
\centering
\caption{Students admission type distribution at UCF over 10 years}
\label{tab:admission type}       
\begin{tabular}{lll}
\hline\noalign{\smallskip}
{}&First-Time-in-College&Transfer\\
\noalign{\smallskip}\hline\noalign{\smallskip}
Percentage&~~~~~~~~~~~39.5\% & ~60.5\% \\
\noalign{\smallskip}\hline
\end{tabular}
\end{table}

\begin{table}
\centering
\caption{Enrollment type distribution for different semesters at UCF over 10 years}
\label{tab:Full-time part time semesters}       
\begin{tabular}{lll}
\hline\noalign{\smallskip}
Semester&Full-time&Part-time\\
\noalign{\smallskip}\hline\noalign{\smallskip}
Fall&~~72.4\% & ~~27.6\% \\
Spring&~~70.6\% & ~~29.4\% \\
Summer&~~10.0\% & ~~90.0\% \\
\noalign{\smallskip}\hline
\end{tabular}
\end{table}

Student record data is processed to extract each student's observed academic load for the semester they enrolled.  Synthetic examples are shown in \tabref{tab:sampleEnrollmentData}. For each student, their enrollment sequence is ordered from their first observed enrollment to their last observed enrollment. The data set includes partial, halted, and graduated enrollment sequences within the indicated ten years date-range.  For the purposes of this study, we restrict the problem to enrollment during Fall and Spring semesters; as such, information regarding Summer enrollment is excluded when constructing the HMM.  It is worth noting that the data-set includes both first-time-in-college students and transfer students.

\section{Applying HMM to student data}
 In applying the HMM model to our problem, we begin by identifying the set of hidden states corresponding to three different enrollment strategies: full-time enrollment strategy (FES), part-time enrollment strategy (PES), and mixed enrollment strategy (MES). The probability of a student changing his enrollment strategy from one semester to the next is represented using the probability transition matrix $A$. The probability of observing a particular enrollment status when a student uses a specific enrollment strategy is given by the emission matrix $B$. Finally, $\pi_0$ is the initial enrollment strategy distribution by which students start their education.

Beginning with an initial guess for $A$, $B$, and $\pi_0$, the Baum-Welch algorithm is applied to estimate the optimal model parameter set, $\lambda^*$. Converging after 20 iterations, the following values for $A$, $B$, and $\pi_o$ are numerically calculated, additionally $\pi$, and the steady-state distribution is provided:
\[
A=\begin{bmatrix}
    0.898~~ & 0.05~~ & 0.052      \\
    0.168~~ & 0.74~~ & 0.092      \\  
    0.007~~ & 0.12~~ & 0.873
\end{bmatrix}
 {},
B=\begin{bmatrix}
    0.974~~ & 0.026\\
    0.611~~ & 0.389\\
    0.061~~ & 0.939
\end{bmatrix} 
\]
\begin{equation*}\pi_0=\begin{bmatrix} 
0.718~~ & 0.113~~ & 0.169
\end{bmatrix}\end{equation*}
\begin{equation*}\pi=\begin{bmatrix} 
0.417~~ & 0.239~~ & 0.344
\end{bmatrix}\end{equation*}

Between any two subsequent semesters at time-steps, $t$ and $t+1$, the rows in the transition matrix $A$ correspond to states FES, MES, and PES at semester $t$, while the columns correspond to states FES, MES, and PES at semester $t+1$. Each element of $A$ thereby describes the probability of switching from one enrollment strategy to another between semesters. Based on the estimated transition matrix $A$, most students maintain their enrollment strategy with high probabilities from one semester to the next.  Reading the matrix's diagonal, with 0.898 probability, a student employing a FES will continue employing a FES. Similarly, with probabilities 0.74 for PES and 0.873 for MES, students maintain their strategy. This finding indicates that most students persist in their enrollment strategy regardless of the strategy they deploy.

For emission matrix $B$, each row corresponds to the probability of full-time and part-time enrollment status in a semester for a given enrollment strategy.  The results indicate that students employing FES register full-time with probability 0.974, conversely, as part-time with probability 0.026. Students employing a PES only register full-time with a probability of 0.061 versus part-time at 0.939. Most interesting are students with MES, as the probability of their full-time and part-time enrollment is split between 0.611 and 0.389. The initial probabilities vector $\pi$ indicates that most undergraduate students start their academic career with a full-time enrollment strategy (with probability 0.718). The initial probabilities for using PES and MES are 0.169 and 0.113, respectively. Furthermore, the steady-state probabilities vector $\pi_0$ shows that  0.417 of students employ a full-time enrollment strategy at any given time. Moreover, the probability of using PES and MES is 0.239 and 0.344, respectively. Comparing $\pi_0$ vs. $\pi$ depicts that although most students have full-time enrollment strategy initially, they are more likely to employ MES and PES for their later semesters than the earlier.  

\section{Demographics Clustering Analysis}
\begin{figure} 
\centering
  \includegraphics[width=0.70\textwidth]{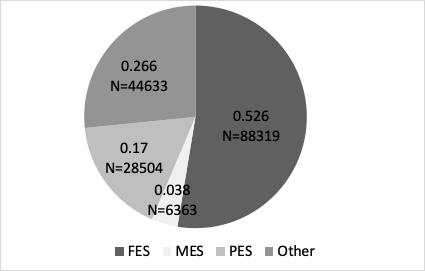}
\caption{Distribution students' enrollment strategy}
\label{fig:Strategy_Dist}
\end{figure}

After estimating the model parameters, the next step is to find the strategies (hidden states) for each student in the data set each semester. Based on the estimated hidden states, students are classified into four groups: three groups corresponding to students who maintain a consistent strategy of FES, PES, or MES during their education, and a fourth group corresponding to students who employ a combination of FES, MES and, PES over their academic career. \figref{fig:Strategy_Dist} shows the distribution of students among these four groups.

Based on results illustrated in \figref{fig:Strategy_Dist}, most students maintain their enrollment strategy during their educational career (the sum of FES, MES, and PES is approximately 73.4\%). The most prevalent consistent enrollment strategy is FES, followed by PES and MES. For students who change strategy at some point during their academic career, 73\% change from FES to PES, and 15\% move from PES to MES. Also, \tabref{tab:comparison UCF with other universities} compares students' enrollment distribution with conducted research \citep{center2017even} among 253 academic institutions. It shows that students at UCF are more likely to register consistently full-time and less likely to enroll part-time than the other universities (35\% vs. 29\% for full-time and 7\% vs. 18\% for part-time).   

\begin{table}
\caption{Distribution over enrollment status and enrollment strategy for UCF and other universities}
\centering
\label{tab:comparison UCF with other universities} 
\begin{tabular}{lllllll}
\hline\noalign{\smallskip}
Target &Always FT&Always PT&FES&MES&PES&Other\\
\noalign{\smallskip}\hline\noalign{\smallskip}
UCF& 35\%&7\%&53\%&3\%&17\%&27\% \\
Prior research& 29\%&18\%&$-$&$-$&$-$&$-$\\
\noalign{\smallskip}\hline
\end{tabular}
\end{table}

\tabref{tab:Gender_strategy} provides the distribution of different enrollment strategies employed by male and female students.
A Chi-square hypothesis test is conducted to see if students' enrollment strategies vary by gender at UCF. The p-value of the Chi-square test (p-value=$2.2e^{-16}$) indicates that the enrollment strategy distribution for female students differs from male students. For example, female students are more likely to adopt a full-time enrollment strategy compared to male students. On the other hand, male students are more likely to apply a combination of enrollment strategies (as illustrated by 'other' in \figref{fig:Strategy_Dist}). \tabref{tab:FTIC_GENDER} and \tabref{tab:TR_GENDER} show gender distribution over the enrollment strategy for FTIC and transfer students separately. Based on these tables, for both FTIC and transfer students, female students with higher probability have the full-time enrollment strategy compared to male students (p-value=$1.18e^{-23}$ for FTIC and p-value=$4.8e^{-14}$ for transfer). However, male FTIC and transfer students are more likely to have the ''other'' enrollment strategy than female students (p-value=$3.88e^{-21}$ for FTIC and p-value=$6.12e^{-15}$ for transfer).

\begin{table}[htbp]
\caption{Female and male ratios for students with different enrollment strategies}
\centering
\label{tab:Gender_strategy}       
\begin{tabular}{llllll}

\hline\noalign{\smallskip}
Gender&FES&MES&PES&Other&Population size\\
\noalign{\smallskip}\hline\noalign{\smallskip}
Male& 52.9\%&3.9\%&14.5\%&28.7\%&~~~~~~62157 \\
Female& 56.4\%&4.0\%&15.5\%&24.1\%&~~~~~~68135\\
\noalign{\smallskip}\hline
\end{tabular}
\end{table}

\begin{table}[htbp]
\centering
\caption{Female and male ratios for FTIC students with different enrollment strategies}
\label{tab:FTIC_GENDER}       
\begin{tabular}{llllll}
\hline\noalign{\smallskip}
Gender&FES&MES&PES&Other&Population size\\
\noalign{\smallskip}\hline\noalign{\smallskip}
Male& 70.8\%&1.1\%&1.2\%&27.0\%&~~~~~~26376 \\
Female& 77.6\%&1.1\%&1.1\%&20.2\%&~~~~~~26748\\
\noalign{\smallskip}\hline
\end{tabular}
\end{table}

\begin{table}[htbp]
\centering
\caption{Female and male ratios for transfer students with different enrollment strategies}
\label{tab:TR_GENDER}       
\begin{tabular}{llllll}
\hline\noalign{\smallskip}
Gender&FES&MES&PES&Other&Population size\\
\noalign{\smallskip}\hline\noalign{\smallskip}
Male& 39.9\%&5.9\%&24.1\%&30.1\%&~~~~~~35431 \\
Female& 42.9\%&5.9\%&24.5\%&26.7\%&~~~~~~41043 \\
\noalign{\smallskip}\hline
\end{tabular}
\end{table}

\tabref{tab:Ethnicity_strategy} describes how enrollment strategies are distributed according to student race/ethnicity. As the table indicates, students who identified as Black and students who identified as Hispanic used FES at similar rates, while students who identified as white used FES at a higher rate than both groups. A hypothesis chi-square post hoc test was conducted to assess if different ethnicity groups use strategies with different distributions. The p-values of all the tests between possible pairs of race/ethnicity groups are close to 0, implying that students with different races/ethnicity have different distributions over the enrollment strategies. \tabref{tab:FTIC_RACE} and \tabref{tab:TR_RACE} summarized this information for FTIC and transfer students, respectively. For both FTIC and transfer students, the enrollment strategy differs between different ethnicity. For example, for transfer students, white students are more likely to employ FES (43\%) than Hispanic students (38.9\%).

\begin{table}[htbp]
\centering
\caption{Ethnicity ratios for students with different enrollment strategies}
\label{tab:Ethnicity_strategy}       
\begin{tabular}{llllll}
\hline\noalign{\smallskip}
Ethnicity&FES&MES&PES&Other&Population size\\
\noalign{\smallskip}\hline\noalign{\smallskip}
White & 56.5\%&3.5\%&13.6\%&26.4\%&~~~~~~71852 \\
Hispanic & 52.0\%&4.5\%&17.4\%&26.1\%&~~~~~~30843 \\
Black & 52.9\%&4.4\%&17.8\%&24.9\%&~~~~~~14545 \\
Other race & 53.8\%&4.1\%&14.5\%&27.6\%&~~~~~~13210 \\
\noalign{\smallskip}\hline
\end{tabular}
\end{table}

\begin{table}[htbp]
\centering
\caption{Ethnicity ratios for FTIC students with different enrollment strategies}
\label{tab:FTIC_RACE}       
\begin{tabular}{llllll}
\hline\noalign{\smallskip}
Ethnicity&FES&MES&PES&Other&Population size\\
\noalign{\smallskip}\hline\noalign{\smallskip}
White & 74.1\%&1.0\%&1.0\%&23.9\%&~~~~~~31208 \\
Hispanic & 74.8\%&1.1\%&1.4\%&22.7\%&~~~~~~11322 \\
Black & 74.9\%&1.5\%&1.1\%&22.5\%&~~~~~~4963 \\
Other race & 72.4\%&1.5\%&1.5\%&24.6\%&~~~~~~5632 \\
\noalign{\smallskip}\hline
\end{tabular}
\end{table}

\begin{table}[htbp]
\centering
\caption{Ethnicity ratios for transfer students with different enrollment strategies}
\label{tab:TR_RACE}       
\begin{tabular}{llllll}
\hline\noalign{\smallskip}
Ethnicity&FES&MES&PES&Other&Population size\\
\noalign{\smallskip}\hline\noalign{\smallskip}
White & 43.0\%&5.6\%&23.0\%&28.4\%&~~~~~~40139 \\
Hispanic & 38.9\%&6.5\%&26.4\%&28.2\%&~~~~~~1939 \\
Black & 41.6\%&5.9\%&26.3\%&26.2\%&~~~~~~9514 \\
Other race & 40.3\%&6.1\%&23.5\%&30.1\%&~~~~~~7424 \\
\noalign{\smallskip}\hline
\end{tabular}
\end{table}

\textbf{Key Finding 1: Both FTIC and Transfer Students:} Students with different genders and races have different enrollment strategies distributions.

\begin{figure}
\centering
  \includegraphics[width=0.75\textwidth]{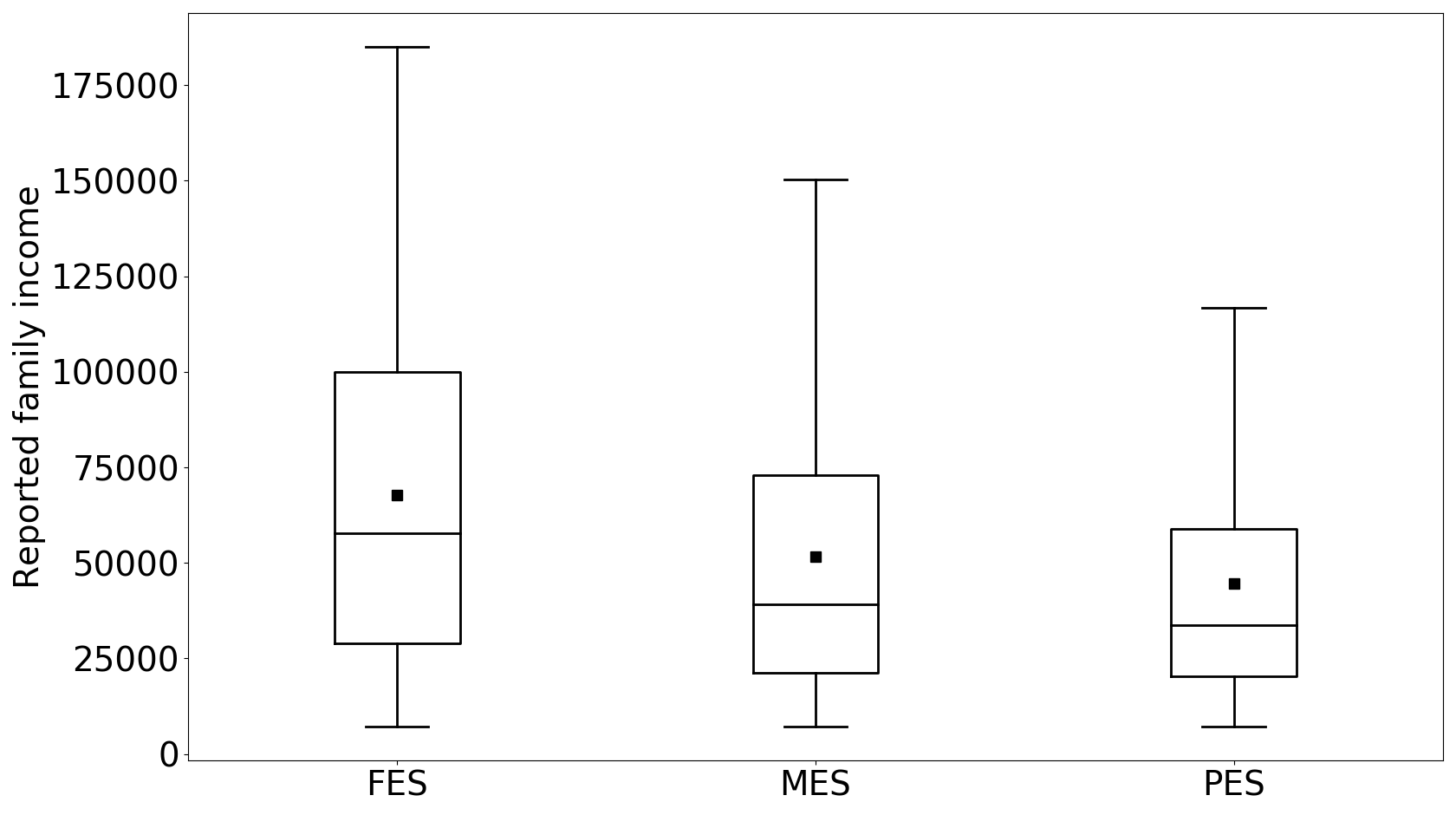}
\caption{Average annually family income for different enrollment strategies, with 5th, 25th, 50th, 75th, and 95th percentiles}
\label{fig:Income}
\end{figure}

The impact of the family financial status on student enrollment strategy in each group is compared in \figref{fig:Income}. Based on the result, students with lower reported family income are more likely to have a part-time enrollment strategy than students with high family income. The Kruskal-Wallis H test was applied to assess if there is a statistically significant difference in family income distribution for students with different enrollment strategies. The p-values associated with the hypothesis test results are zero \footnote{The sample size for FES, MES, and PES is 59400, 4170, and 16371, respectively.}, indicating a significant difference in annual family income distribution between students with different enrollment strategies. While not demonstrated as verified, in this study, we presume that family income or, more specifically, student finance is a crucial factor in students' persistence, as supported by previous studies \citep{davis2011factors}. 

\textbf{Key Finding 2: Financial support:} PES students come from families with lower income levels compared to FES and MES. These students need more financial supports.


\section{Academic Performance Clustering Analysis}
Following students' clustering based on enrollment strategy (FES, PES, MES), several descriptive statistics are calculated. They include average cumulative GPA, graduation rate, and DFW rate.
\subsection{GPA Analysis}
\begin{figure}
\centering
  \includegraphics[width=0.75\textwidth]{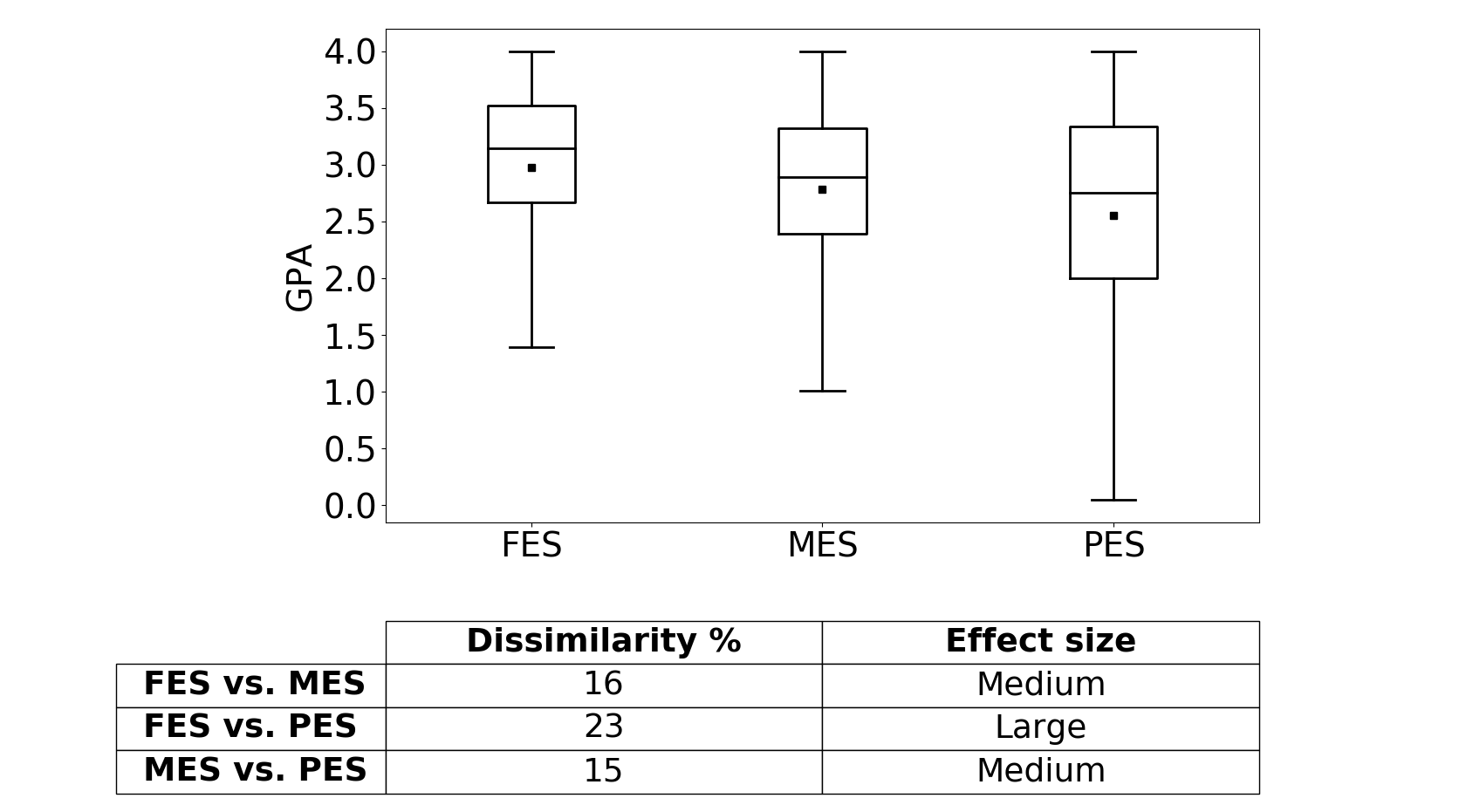}
\caption{Average GPA for different enrollment strategies, with 5th, 25th, 50th, 75th, and 95th percentiles}
\label{fig:Ave_GPA}
\end{figure}

The average GPA\footnote{For each student, first, we compute the average cumulative GPA over his semesters, and then for each enrollment strategy group, we take the average over students' GPA in the group.} for each strategy cluster is shown in \figref{fig:Ave_GPA}. Results show that the FES group has the highest average GPA (2.97). The lowest GPA corresponds to the PES group (2.55), while the MES group's GPA (2.78) lies between PES and FES. Statistical hypothesis Games-Howell post hoc multiple comparison tests are conducted to assess if the average GPA for each group is statistically different from the average GPA of other groups \footnote{Since our samples do not have the same variance, we cannot run the ANOVA test. We use the Games-Howell test that is non-parametric and does not assume homogeneity of variances.}. The result shows that the p-values for all the hypothesis tests are 0, indicating that each group's average GPA is statistically different from others  \footnote{The sample size for FES, MES, and PES is 88319, 6363, and 28504, respectively.}. The table in \figref{fig:Ave_GPA} shows the effect size and \emph{dissimilarity\%}, that is the percentage by which the distribution of each group differs from each other. The relationship between dissimilarity\% and Cohen's d metric is explained in Appendix A. Based on the table, the difference in GPA between FES vs. PES groups is more significant than the other pairs. Furthermore, \figref{fig:Ave_GPA} indicates that students who employed FES and MES had more consistent performance than those who used PES with a wider GPA percentile. These findings suggest that increased student engagement with the university is associated with higher academic performance. Therefore university policymakers may help students improve their academic performance by supporting them to increase their engagement with the university.

\begin{figure}
\centering
  \includegraphics[width=0.75\textwidth]{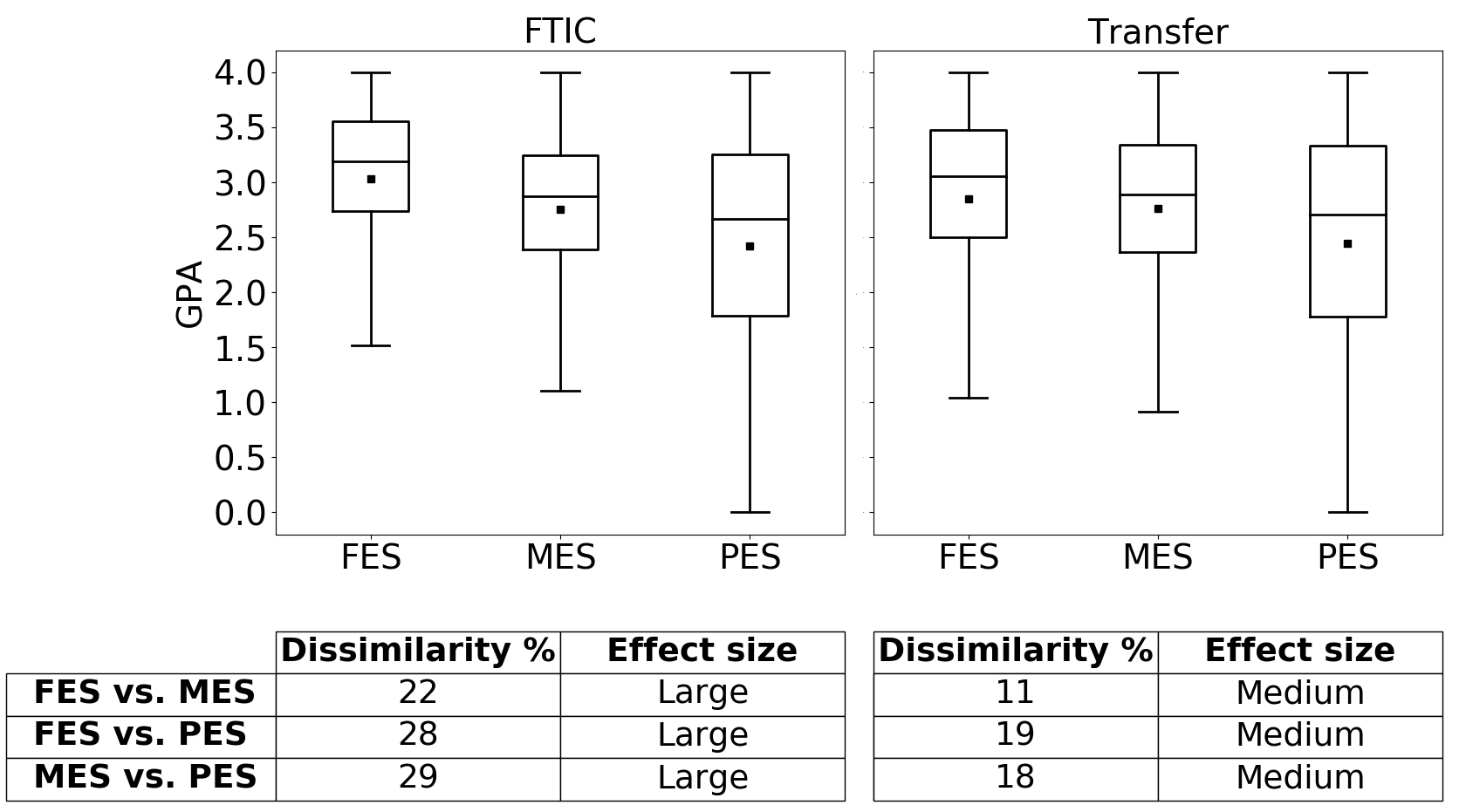}
\caption{Average GPA for FTIC and transfer students with different enrollment strategies, with 5th, 25th, 50th, 75th, and 95th percentiles}
\label{fig:FTIC_TR_GPA}
\end{figure}

\figref{fig:FTIC_TR_GPA} illustrates the average GPA for FES, MES, and PES groups separately for FTIC and transfer students. The average GPA for FTIC and transfer students are 2.99 and 2.75, respectively. For both FTIC and transfer students, FES groups have the highest average GPAs (3.03 for FTIC and 2.85 for transfer), followed by MES (2.75 for FTIC and 2.76 for transfer), and PES (2.42 for FTIC and 2.44 for transfer) respectively. Also, PES students have a less consistent GPA (wider interquartile range) compared to FES and MES students for both groups. The effect size table in the figure shows that the mean GPA differences between different enrollment strategies for FTIC students (large) are larger than the difference for transfer students (medium). In fact, FTIC students' GPAs are more sensitive to enrollment strategy compared to transfer students' GPAs.

Conducted KS2 tests show that for both FTIC and transfer students, the GPA distribution for FES, MES, and PES groups statically differ from each other.\footnote{For FTIC students, sample sizes for FES, MES, and PES groups are 39414, 578, and 600, respectively. For transfer students, these sample sizes are 31748, 4515, 18580, respectively.} Also, we compared the mean GPA between FTIC and transfer students with different enrollment strategies. For FES groups, FTIC students have a higher mean GPA compared to transfer students, and the difference is statistically significant (with small effect size). On the other hand, for MES and PES groups, there is no difference in mean GPAs between FTIC and transfer students \footnote{Games-Howell post hoc multiple comparison tests are conducted to see if the mean of each group differs from the other groups.}. In other words, for MES and PES groups, transfer students have a similar performance to FTIC students.

\textbf{Key Finding 3: Both FTIC and Transfer Students:} FES students have the highest GPA, followed by MES, followed by PES. The higher the student's engagement, the higher is the student's GPA.



\begin{figure}
\centering
  \includegraphics[width=0.75\textwidth]{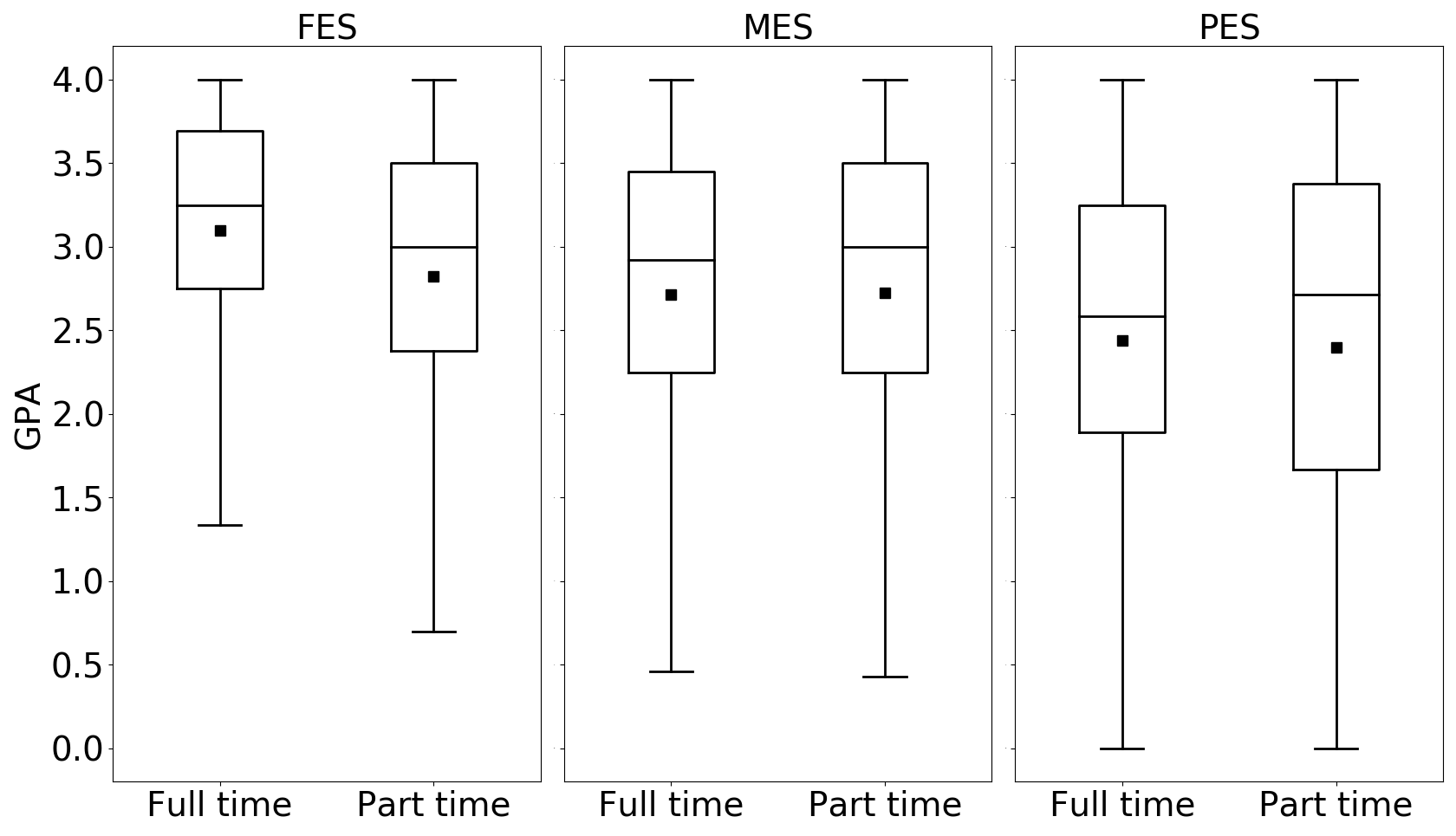}
\caption{Average GPA for FTIC students with different enrollment strategies, with 5th, 25th, 50th, 75th, and 95th percentiles}
\label{fig:FTIC_GPA_SEMESTERS}
\end{figure}

Furthermore, inside each strategy cluster, the average GPA during full-time and part-time semesters are calculated. As indicated in \figref{fig:FTIC_GPA_SEMESTERS}, FTIC students who adopt a full-time enrollment strategy have a higher average GPA in full-time semesters than their part-time semesters (3.1 vs. 2.82). This finding indicates that students employing a full-time enrollment strategy tend not to perform well when registering part-time. However, of interest is that for students using a mixed-enrollment strategy, hypothesis tests indicate there is no statistical difference in the means between the GPAs of full-time and part-time semesters. In other words, the results suggest that semester enrollment status for MES students does not significantly impact their GPAs \footnote{The sample size for full-time and part-time semesters are 1576 and 1281, respectively.}. While the GPA reductions observed for students employing a full-time enrollment strategy appear reasonable, the lack of GPA drop for mixed enrollment strategy students is somewhat surprising. It suggests potential value in encouraging part-time students to occasionally enroll full-time. The same conclusion is observed for students in the PES group; that is, there is no statistical difference between the average GPAs of full-time and part-time semesters \footnote{The sample size for full-time and part-time semesters are 107 and 1556, respectively.}.

\begin{figure}
\centering
  \includegraphics[width=0.75\textwidth]{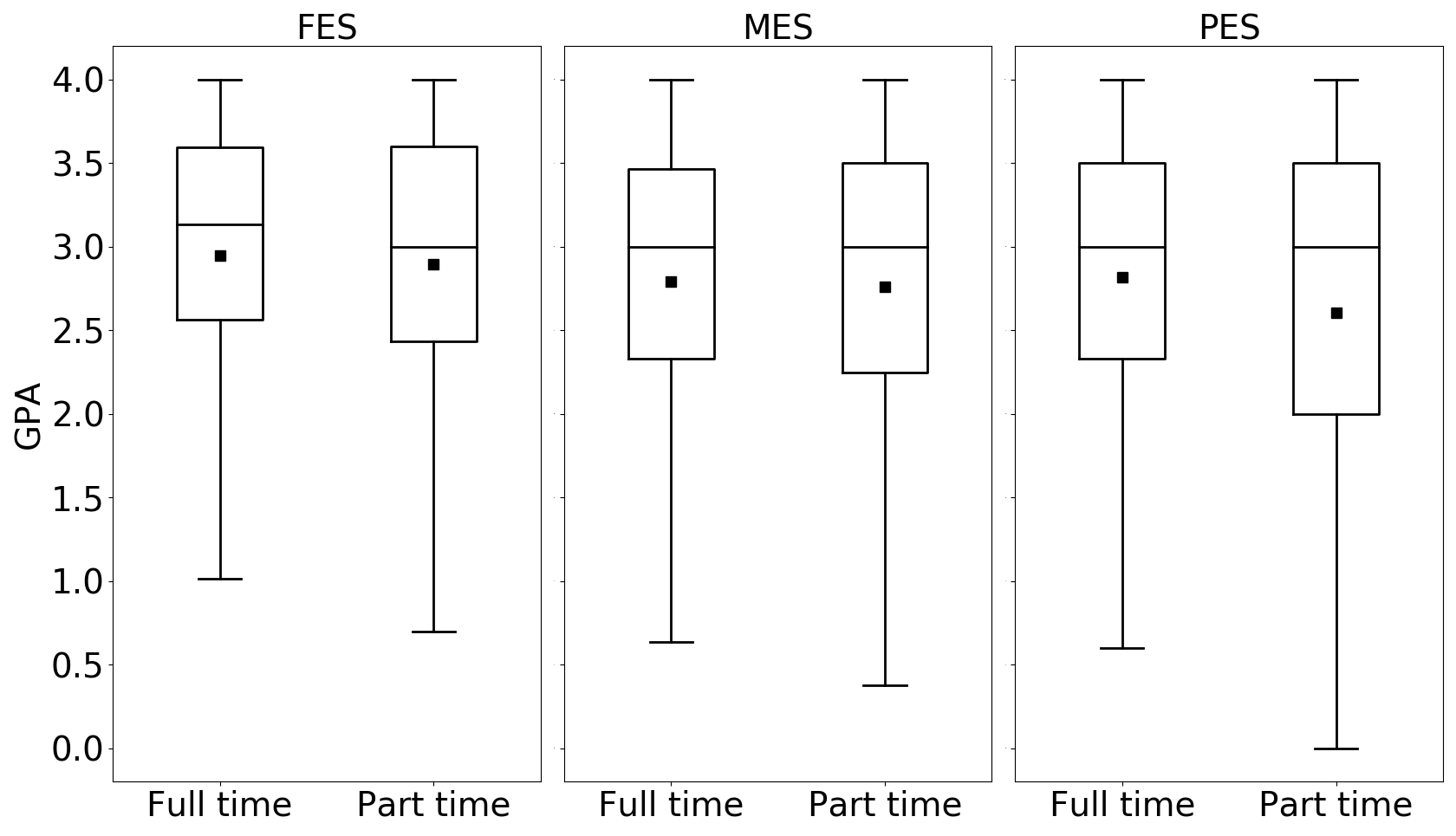}
\caption{Average GPA for transfer students with different enrollment strategies, with 5th, 25th, 50th, 75th, and 95th percentiles}
\label{fig:TR_GPA_SEMESTERS}
\end{figure}

In \figref{fig:TR_GPA_SEMESTERS}, we perform a similar analysis focusing on transfer students. The results indicated that FES and PES students have a higher GPA when registered full-time than when registered part-time (2.95$>$2.89 and 2.81$>$2.6). However, there is no statistical difference between GPA for full-time and part-time semesters for the MES group \footnote{The sample size for full-time and part-time semesters are 10480 and 9390, respectively}. \tabref{tab:FTIC effect size} and \tabref{tab:Tr effect size} show the computed effect size between GPA distributions in full-time and part-time semesters for FTIC and transfer students. As we see in these tables, the effect sizes for FTIC students (\tabref{tab:FTIC effect size}) are larger than transfer students (\tabref{tab:Tr effect size}), which implies transfer students are less sensitive to enrollment strategy decisions compared to FTIC students. 

Also, some comparisons are conducted between FTIC and transfer students in terms of average GPA in full-time and part-time semesters. For example, for the PES group, the average GPA for transfer students in full-time semesters is 2.81, while for FTIC students, it is 2.44. These values for part-time semesters are 2.60 and 2.40, and the differences are statistically significant. In other words, PES transfer students have a higher GPA compared to PES FTIC students for their full-time and part-time semesters. These results may seem counter-intuitive while it can be explained by the academic history of transfer students. These students who usually transfer from community collages are used to take part-time jobs and, therefore, part-time semesters, which justifies them maintaining reasonable academic performance in part-time semesters \citep{fredrickson1998today}.

\begin{table}
\begin{minipage}{1.0\textwidth}
\caption{Effect size and distribution dissimilarity percentage between full-time and part-time semester GPA for FTIC students with different enrollment strategies }
\begin{tabular}{lllll}
\hline\noalign{\smallskip}
~~~~~Pairs & Semester & Dissimilarity \% & Effect size \\
\noalign{\smallskip}\hline\noalign{\smallskip}
\multirow{2}{*}{FES vs. MES}
 & Full-time  &~~~~~~~~19\% &~~Medium\\
 & Part-time  &~~~~~~~~5\% &~~Small\\\hline
\multirow{2}{*}{FES vs. PES}
 & Full-time  &~~~~~~~~36\% &~~Very Large\\
 & Part-time  &~~~~~~~~17\% &~~Medium\\\hline
\multirow{2}{*}{MES vs. PES}
 & Full-time  &~~~~~~~~22\% &~~Large\\
 & Part-time  &~~~~~~~~15\% &~~Medium\\\hline
\end{tabular}
\label{tab:FTIC effect size} 
\end{minipage}%
\hfill
\end{table}

\begin{table}
\centering
\begin{minipage}{1.0\textwidth}
\caption{Effect size and distribution dissimilarity percentage between full-time and part-time semester GPA for transfer students with different enrollment strategies}
\begin{tabular}{llll}
\hline\noalign{\smallskip}
~~~~~Pairs & Semester & Dissimilarity \% & Effect size \\
\noalign{\smallskip}\hline\noalign{\smallskip}
\multirow{2}{*}{FES vs. MES}
 & Full-time  &~~~~~~~~10\% &~~Medium\\
 & Part-time  &~~~~~~~~6\% &~~Small\\\hline
\multirow{2}{*}{FES vs. PES}
 & Full-time  &~~~~~~~~9\% &~~Small\\
 & Part-time  &~~~~~~~~14\% &~~Medium\\\hline
\multirow{2}{*}{MES vs. PES}
 & Full-time  &~~~~~~~~6\% &~~Small\\
 & Part-time  &~~~~~~~~11\% &~~Medium\\\hline
\end{tabular}
\label{tab:Tr effect size} 
\end{minipage}%
\hfill
\end{table}

\textbf{Key Finding 4: FTIC Students:}
FES students, when register as full-time, have a higher GPA compared to when they register part-time. There is no difference between the GPA of the full-time and part-time semester for MES and PES students.

\textbf{Key Finding 5: Transfer Students:}
FES and PES students, when register as full-time, have a higher GPA compared to when they register part-time. For MES students, there is no difference between the GPA of the full-time and part-time semester.

\textbf{Key Finding 6: FTIC and Transfer Students:}

For PES students with full-time and part-time enrollment, transfer students have a higher GPA than FTIC students.

\subsection{DFW rate Analysis}

\begin{figure}
\centering
  \includegraphics[width=0.75\textwidth]{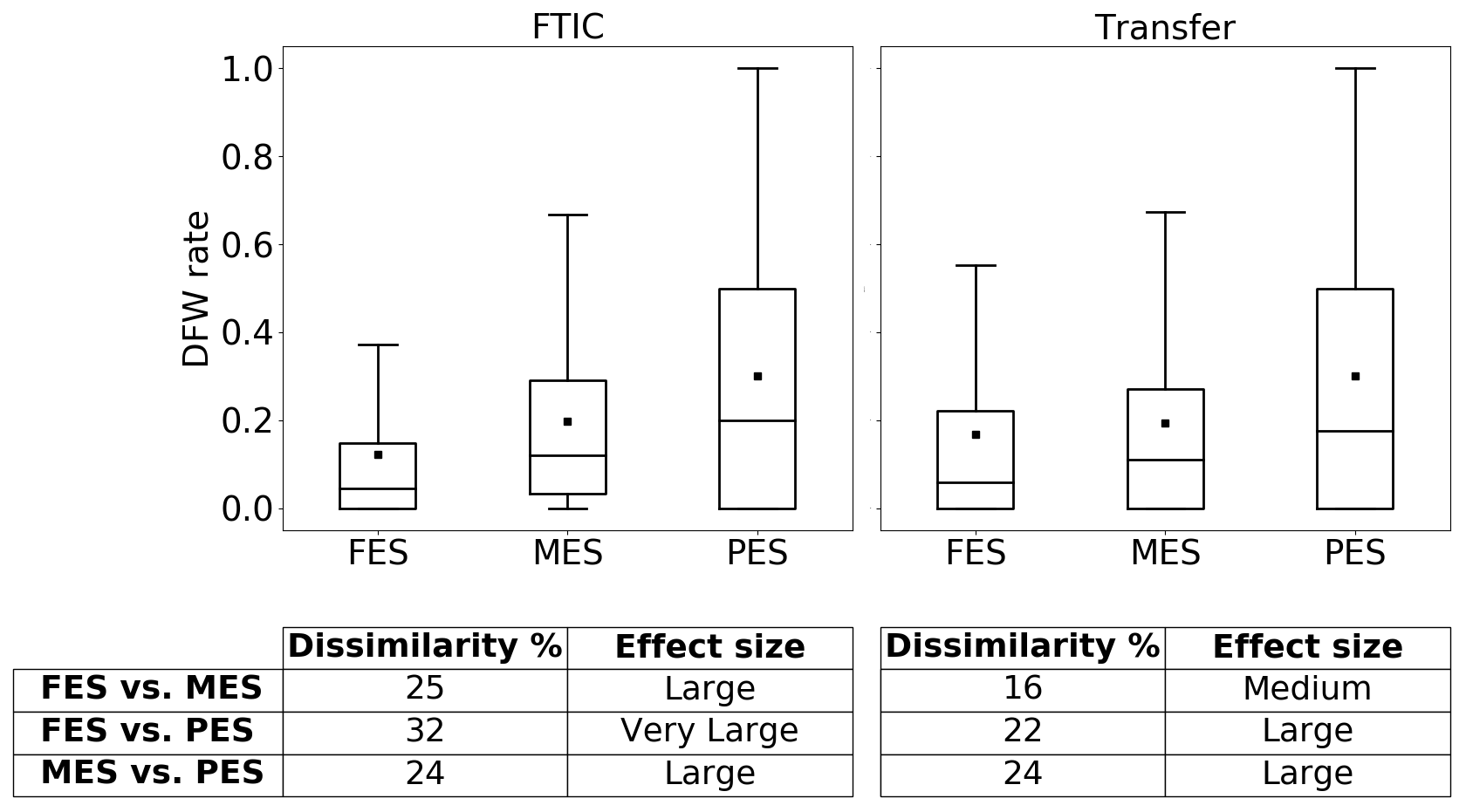}
\caption{Average DFW rate for FTIC and transfer students with different enrollment strategies, with 5th, 25th, 50th, 75th, and 95th percentiles}
\label{fig:FTIC_TR_DFW}
\end{figure}

As another indicator of students' performance, we compare the DFW rate for students with different enrollment strategies. The DFW rate is defined as the number of courses with D, F, and W grades over the total number of courses taken by a student. Appendix B shows an example of computing the DFW rate. Whisker plots for DFW rates separated by FTIC and transfer students are provided in \figref{fig:FTIC_TR_DFW}. As expected, for both FTIC and transfer students, the PES group are most likely to get DFW grades, followed by MES and FES, respectively, and the differences are statistically significant \footnote{Games-Howell post hoc multiple comparison tests are conducted to see if the average DFW rates of each group differ from other groups. For both FTIC and transfer students, the p-values of the tests for all possible pairs between FES, MES, and FES is 0, which means the difference is statistically significant. For FTIC students, the sample size of FES, MES, and PES groups is 39371, 578, and 596, respectively; and for transfer students, these values are 31721, 4515, and 18521, respectively.}. 

\begin{figure}[htbp]
\centering
  \includegraphics[width=0.75\textwidth]{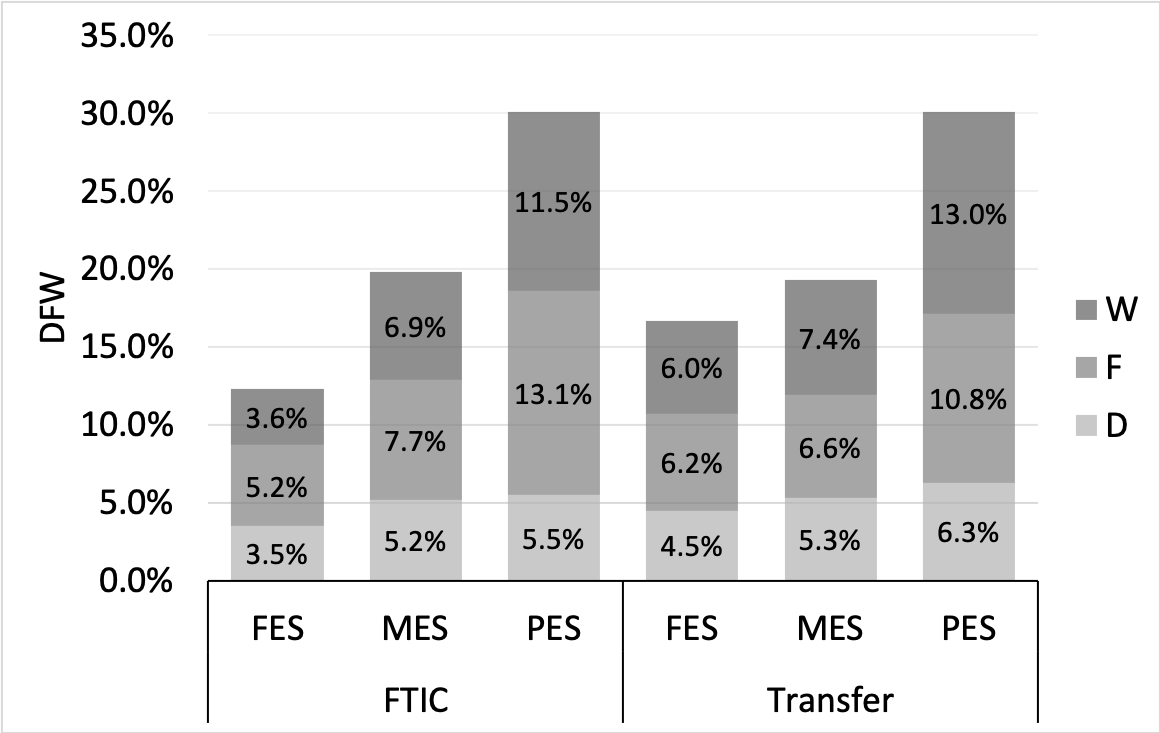}
\caption{Average D, F, and W rate for FTIC and transfer students with different enrollment strategies}
\label{fig:DFW_Split}
\end{figure}

Furthermore, effect size results in the figure show that the mean DFW rate for FTIC students is more sensitive to their enrollment strategy than transfer students. When comparing FTIC and transfer students in terms of the DFW rate, we find that for the FES group, FTIC students have a lower average DFW rate compared to transfer students, and the difference is statistically significant. However, for MES and PES groups, there is no significant difference in the mean DFW rate.
To get a better idea of the grade distribution, \figref{fig:DFW_Split} represents DFW rates broken to D, F, and W for FTIC and transfer students separately. Based on the figure, following the same pattern as the DFW rates, FES students have the lowest rates in each of D, F, and W grades when compared to MES and PES students, who follow respectively.
Moreover, among those students who are not in an overall good standing (having DFW grades), those adopting PES show more intention to drop the course (\%38 FTIC and \%54 transfer) compared to MES (\%35 FTIC and \%38 transfer) and FES (\%29 FTIC and \%36 transfer). 

\textbf{Key Finding 7: FTIC and Transfer Students:}
PES students have a higher DFW rate compared to MES and FES groups.

\subsection{Graduation rate Analysis}
Another important metric in assessing academic performance for different enrollment strategies is the graduation rate. To compute this metric, American universities use the six-year graduation rate approach. Based on the federal definition, the six-year graduation rate refers to the percentage of FTIC students who complete their program within 150\% of the standard enrollment time to degree \citep{hagedorn2005define}. For example, for a four-year program, students who earn degrees within six years are considered graduates.

Despite its popularity, the six-year graduation rate approach has two main blind spots. First, no distinction exists between students with different enrollment strategies. PES students often take fewer courses each semester compared to MES and FES. Therefore, they are more likely to finish their program in more than six years. All such students are considered halting their program in the six-year graduation rate method, leading to underestimating the actual graduation rate. Second, the six-year graduation rate is only computed for FTIC students, excluding transfer students.

In this paper, we introduce two alternative methodologies for computing the graduation rate. In the first methodology, we calculate the halt rate as a substitute for the graduation rate. The halt rate refers to the percentage of FTIC students who halt enrollment at UCF within six years. Students are considered as halt if they do not register at UCF for three consecutive semesters. This new measure helps us avoid the underestimation caused by excluding students who take more than six years to graduate. 

In the second approach, we use absorbing Markov Chain to compute the graduation rates. This approach doesn't exclude students who graduate after six years, as it can calculate the graduation/halt rate probability for any duration of time that students spend in their program. Moreover, it makes it possible to well consider transfer students by inserting them into the Markov chain model, taking into account their academic level when they join the new program. This approach is explained in detail by \citet{boumi2020improving}.  

\tabref{tab:G_H rates} shows the six-year graduation and halt rates for FTIC students who start their program in Fall 2008, 2009, and 2010. As shown in the table, PES students have a higher six-year halt rate than FES and MES students.

\begin{table}
\centering
\caption{six-year graduation and halt rate for FTIC students who start in Fall 2008, 2009, and 2010}
\label{tab:G_H rates}       
\begin{tabular}{lll}
\hline\noalign{\smallskip}
Strategy&Graduation rate&Halt rate\\
\noalign{\smallskip}\hline\noalign{\smallskip}
~~~FES&~~~~~~~~69\% & ~~~30\% \\
~~~MES&~~~~~~~~41\% & ~~~51\% \\
~~~PES&~~~~~~~~16\% & ~~~81\% \\
~~~Other&~~~~~~~~82\% & ~~~17\% \\
\noalign{\smallskip}\hline
\end{tabular}
\end{table}

\begin{table}
\begin{minipage}{1.0\textwidth}
\caption{Graduation rate and time (semesters) to finish school for FTIC students with different enrollment strategies (*Graduation Rate)}
\begin{tabular}{lllll}
\hline\noalign{\smallskip}
Strategy & States & G rate$^*$ & Time to graduate & Time to halt \\
\noalign{\smallskip}\hline\noalign{\smallskip}
\multirow{5}{*}{FES}
 & Start  &~~66\% & ~~~~~~~11.56&~~~~~~4.75\\
 & Freshman  &~~63\% & ~~~~~~~11.09&~~~~~~3.82\\
 &Sophomore &~~78\% & ~~~~~~~8.78&~~~~~~3.29\\
 &Junior & ~~$\mathbf{90\%}$&~~~~~~~6.27&~~~~~~3.16 \\
 &Senior &~~97\% & ~~~~~~~3.64&~~~~~~2.78\\
\hline
\multirow{5}{*}{MES}
 & Start  &~~49\% & ~~~~~~~12.57&~~~~~~5.52\\
 & Freshman  &~~43\% & ~~~~~~~12.74 &~~~~~~4.70\\
 &Sophomore &~~64\%& ~~~~~~~10.20 &~~~~~~4.18\\
 &Junior &~~$\mathbf{77\%}$ & ~~~~~~~7.44 &~~~~~~2.97\\
 &Senior &~~93\% & ~~~~~~~4.58 &~~~~~~3.3\\\hline
 \multirow{5}{*}{PES}
 & Start  &~~12\% & ~~~~~~~12.10&~~~~~~4.00\\
 & Freshman  &~~6\%& ~~~~~~~13.26 &~~~~~~3.15 \\
 &Sophomore &~~15\%& ~~~~~~~11.17 &~~~~~~2.65 \\
 &Junior &~~$\mathbf{38\%}$& ~~~~~~~8.48 &~~~~~~2.58\\
 &Senior &~~82\%& ~~~~~~~5.33 & ~~~~~~2.44\\\hline
 \multirow{5}{*}{Other}
 & Start  &~~80\% & ~~~~~~~13.17&~~~~~~8.32\\
 & Freshman  &~~79\%& ~~~~~~~12.65 &~~~~~~7.58 \\
 &Sophomore &~~82\%& ~~~~~~~10.27 &~~~~~~5.97 \\
 &Junior &~~$\mathbf{88\%}$& ~~~~~~~7.61 &~~~~~~4.79\\
 &Senior &~~94\%& ~~~~~~~4.8 & ~~~~~~4.23\\\hline
\end{tabular}
\label{tab:FTIC_G_rate} 
\end{minipage}%
\hfill
\end{table}

\begin{table}
\begin{minipage}{1.0\textwidth}
\caption{Graduation rate and time (semesters) to finish school for transfer students with different enrollment strategies (*Graduation Rate)}
\begin{tabular}{lllll}
\hline\noalign{\smallskip}
Strategy & States & G rate$^*$ & Time to graduate & Time to halt \\
\noalign{\smallskip}\hline\noalign{\smallskip}
\multirow{5}{*}{FES}
 & Start  & ~~74\%&~~~~~~~7.08&~~~~~~3.68 \\
 & Freshman  & ~~58\%&~~~~~~~8.28&~~~~~~3.01 \\
 &Sophomore & ~~62\%&~~~~~~~7.88&~~~~~~2.74 \\
 &Junior &~~$\mathbf{73\%}$&~~~~~~~6.27&~~~~~~2.69 \\
 &Senior &~~89\%&~~~~~~~3.82&~~~~~~2.51 \\
\hline
\multirow{5}{*}{MES} 
 & Start  & ~~72\%&~~~~~~~8.15&~~~~~~4.74 \\
 & Freshman  &~~67\%&~~~~~~~8.41&~~~~~~4.19 \\
 &Sophomore &~~64\% &~~~~~~~8.85&~~~~~~4.05\\
 &Junior &~~$\mathbf{71\%}$&~~~~~~~7.51&~~~~~~3.78 \\
 &Senior &~~85\% &~~~~~~~4.78&~~~~~~3.41 \\\hline
 \multirow{5}{*}{PES}
 & Start  & ~~36\%&~~~~~~~9.5&~~~~~~3.73 \\
 & Freshman  &~~26\% &~~~~~~~9.36&~~~~~~2.76 \\
 &Sophomore &~~17\% &~~~~~~~10.25&~~~~~~2.36 \\
 &Junior &~~$\mathbf{29\%}$ &~~~~~~~9.3&~~~~~~2.73 \\
 &Senior &~~61\% &~~~~~~~6.13&~~~~~~2.79 \\\hline
 \multirow{5}{*}{Other}
 & Start  & ~~82\%&~~~~~~~8.74&~~~~~~5.94 \\
 & Freshman  &~~79\% &~~~~~~~9.59&~~~~~~6.14 \\
 &Sophomore &~~78\% &~~~~~~~9.49&~~~~~~5.55 \\
 &Junior &~~$\mathbf{81\%}$ &~~~~~~~8.05&~~~~~~5.01 \\
 &Senior &~~88\% &~~~~~~~5.36&~~~~~~4.21 \\\hline
\end{tabular}
\label{tab:TR_G_rate} 
\end{minipage}%
\hfill
\end{table}

\tabref{tab:FTIC_G_rate} shows the graduation rate for FTIC students with different types of enrollment strategies using the absorbing Markov chain method. As we see in the table, FES students have the highest graduation rate for any academic level, followed by MES and PES, respectively. Furthermore, for FTIC students with any enrollment strategy, higher academic levels correspond to higher graduation rates. For example, Senior students have a higher graduation rate compared to Junior, Sophomore, and Freshman students.
\begin{figure}[h!]
\centering
  \includegraphics[width=0.65\textwidth]{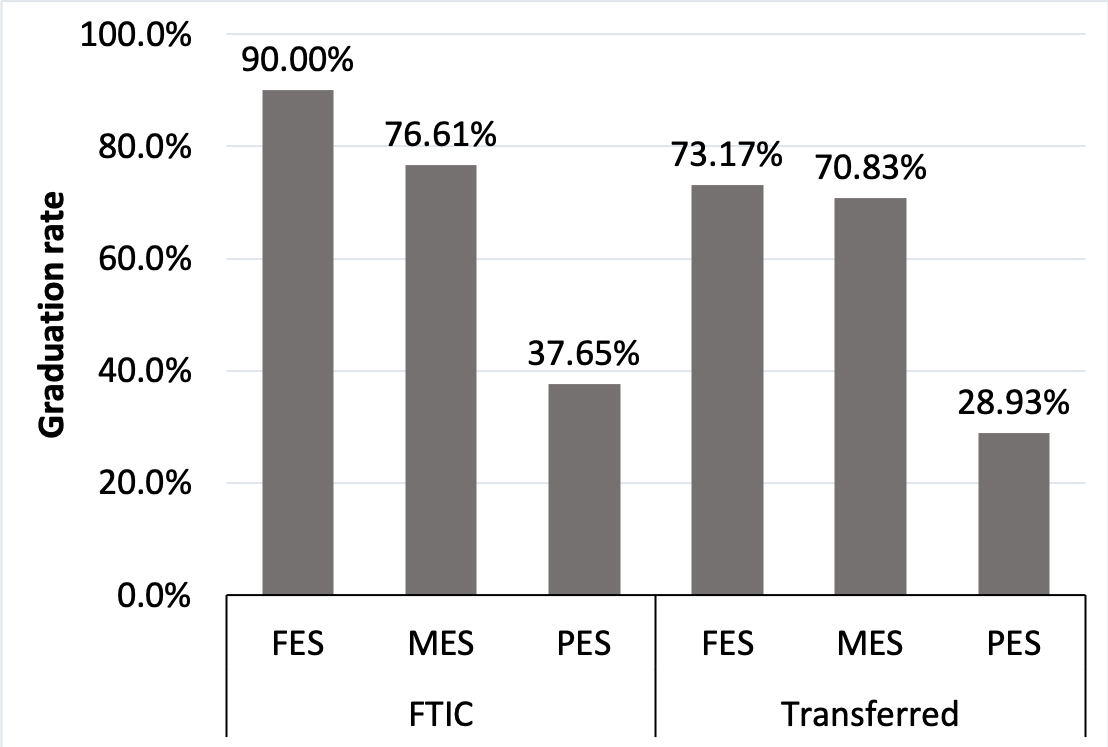}
\caption{Graduation rate for FTIC and transfer students with junior academic level and different enrollment strategies}
\label{fig:G_Rate}
\end{figure}

Columns \emph{Time to graduate} and  \emph{Time to halt} in \tabref{tab:FTIC_G_rate} explain expected time that takes for students to graduate and halt enrollment respectively. As shown in the table, the time to graduate for the FES group is notably shorter than for MES and PES groups. For example, students in the FES group with \emph{Freshman} academic level need on average of 11.09 semesters to graduate from their program, while for students in MES and PES groups, it takes 12.74 and 13.26 semesters, respectively. On the other hand, we see a different order in time to halt for different enrollment strategies. As an example, time to halt for freshman students in MES (4.70) group is longer than FES (3.82) and PES (3.15) groups. This could be explained as some full-time students may decide to alternate between full-time and part-time semesters as a strategy to decrease imposed workload and avoid halting enrollment. \tabref{tab:TR_G_rate} represents the same measures in \tabref{tab:FTIC_G_rate} for  transfer students. For both FTIC and transfer groups, FES students have the highest graduation rate, followed by MES, and PES, respectively. These results are shown in \figref{fig:G_Rate}.

\textbf{Key Finding 7: FTIC and Transfer Students:}

FES students have a higher graduation rate than MES and PES groups. However, MES students have a high graduation rate as well.

\textbf{Key Finding 9: Transfer Students:}
Time to graduate for PES students is greater than time to graduate for MES and FES students. MES is a good strategy for transfer students since graduate rate for this group is as high as FES group.

\textbf{Key Finding 10: FTIC Students:}
MES and PES students almost have the same time to graduate, which is greater than the time to graduate for FES students.
%
\section{Discussion}
In section 8, we showed that students who are more engaged with their study program have a better academic performance on average. In this section, we use the \emph{other} group of students -  students who use a combination of the FES, MES, and PES during their academic career - to validate our results.

The graduation rate results in the previous section showed that PES students are more likely to halt school before graduation. This finding could reasonably raise concerns over the possibility of dropouts for those students who change their enrollment strategy from FES to PES. To assess this, we compare enrollment persistence of students who switch from FES to PES with those who maintain FES. To remove any biased source, we consider students in the same college and have similar GPAs. Results indicated that around \%16 of students in the college of science, who switched after three FES semesters to PES, have dropped out from their program. This percentage is significantly lower for students who maintain FES throughout the same period (less than 1\% dropout). Results for other colleges are summarized in \tabref{tab:Dropout}.

\begin{table}
\centering
\caption{Dropout ratio comparison between students who stay FES and students who switch from FES to PES for different colleges }
\label{tab:Dropout}       
\begin{tabular}{lll}
\hline\noalign{\smallskip}
College&FES to PES students&staying FES students\\
\noalign{\smallskip}\hline\noalign{\smallskip}
Science & 10 out of 62& 21 out of 3246 \\
Engr \& Comp Sci & 3 out of 26&13 out of 1789 \\
Medicine & 1 out of 4& 5 out of 651 \\
Business & 2 out of 35&11 out of 1506 \\
\noalign{\smallskip}\hline
\end{tabular}
\end{table}

\begin{figure}
\centering
  \includegraphics[width=0.75\textwidth]{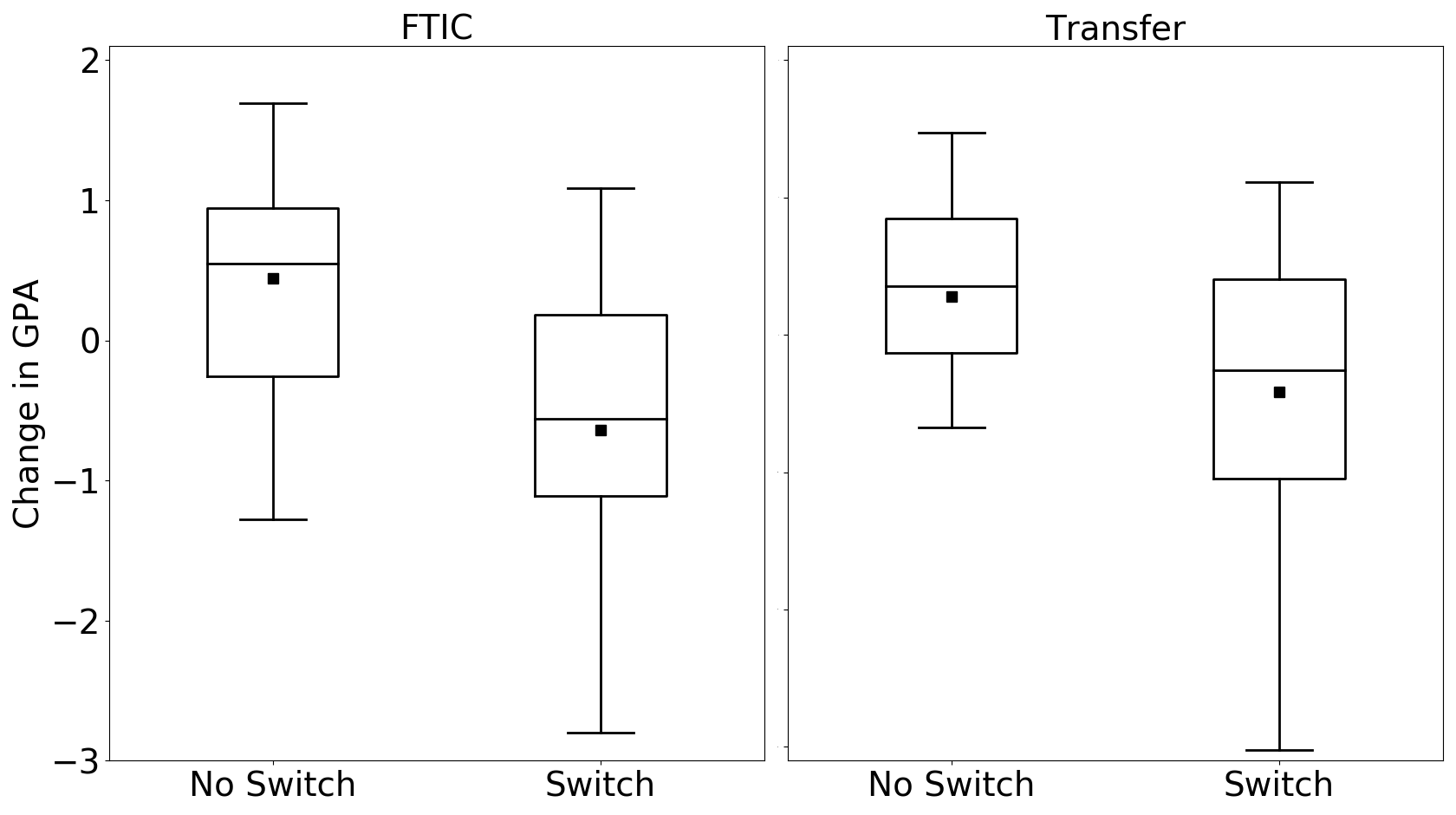}
\caption{comparing changes in GPA between students who switch from FES to PES and students who stay FES for FTIC and transfer students}
\label{fig:FTIC_TR_FES_TO_PES_change_GPA}
\end{figure}

Next, we compare the GPA trend between students who switch their enrollment strategy to PES after three FES semesters and those who stay FES through their academic career. Again, students in both groups are selected from the same colleges and have similar GPAs at the end of the third semester. To get the GPA trend, we compute the difference between the fourth semester GPA and the average GPA for the first three semesters and compare the result between the two groups. For FTIC students, the average difference for students who switch from FES to PES is -0.64, while for students who stay FES is +0.44. These results for transfer students are +0.28 and -0.42, respectively. More detailed statistics are provided in  \figref{fig:FTIC_TR_FES_TO_PES_change_GPA} for FTIC and transfer students separately.

We use two different statistical tests to examine if the decrease in the GPA for those who have switched their enrollment strategy from FES to PES is significant: (1) the hypothesis t-test, and (2) the two-ways Anova test. The P-value for the hypotheses t-tests (0) shows that for both FTIC and transfer students, the difference between the average GPA of the three first semesters and the fourth semester is statistically significant. Although this could prove quite informative, one may argue that different syllabus and/or educational policies in different colleges can impact students' behavior. 

We use a two-ways ANOVA test to investigate the impact of switching enrollment strategy together with the college of enrollment on the observed decline in GPA. Therefore, a linear regression model is fitted on data with two independent variables; one categorical variable, which represents the college of enrollment, and one binary variable, which indicates if a switch is made from FES to PES after three semesters. The switching variable is equal to 1 if the student switches from FES to PES in the fourth semester, and is equal to 0 otherwise. We consider the difference between the fourth semester GPA and average GPA for the three first FES semesters as the response variable. The results of ANOVA test for FTIC and transfer students are summarized in \tabref{tab:ANOVA_FTIC_FES_To_PES} and \tabref{tab:ANOVA_TR_FES_To_PES} respectively. 

As we see in these tables, the switching variable (with the p-value close to 0) is the only factor impacting students' GPA. This finding implies that regardless of enrollment colleges, such changes in students' enrollment strategy are crucial to be understood and monitored and can help university policymakers decrease chances for students' academic failure by taking preventive actions on time.

\begin{table}
\centering
\caption{Analysis of variance for the linear regression model for FTIC students switching from FES to PES}
\label{tab:ANOVA_FTIC_FES_To_PES}       
\begin{tabular}{llllll}
\hline\noalign{\smallskip}
Source&DF&Adj SS&Adj MS&F-Value&P-Value\\
\noalign{\smallskip}\hline\noalign{\smallskip}
College &7 &6.388&0.9125&1.11&0.367  \\
Switching &1 &19.995&19.9952&24.41&0.000 \\
College*Switching &7 &9.219&1.317&1.61&0.152  \\
Error &56 &45.875&0.8192&  &   \\
Total &71 &82.442 &  & &   \\
\noalign{\smallskip}\hline
\end{tabular}
\end{table}

\begin{table}
\centering
\caption{Analysis of variance for the linear regression model for transfer students switching from FES to PES }
\label{tab:ANOVA_TR_FES_To_PES}       
\begin{tabular}{llllll}
\hline\noalign{\smallskip}
Source&DF&Adj SS&Adj MS&F-Value&P-Value\\
\noalign{\smallskip}\hline\noalign{\smallskip}
College &7 &3.548&0.5068&0.55&0.797  \\
Switching &1 &9.485&9.4854&10.2&0.002 \\
College*Switching &7 &1.747&0.2496&0.27&0.964  \\
Error &62 &57.639&0.9297&  &   \\
Total &77 &72.420 &  & &   \\
\noalign{\smallskip}\hline
\end{tabular}
\end{table}

\begin{figure}
\centering
  \includegraphics[width=0.75\textwidth]{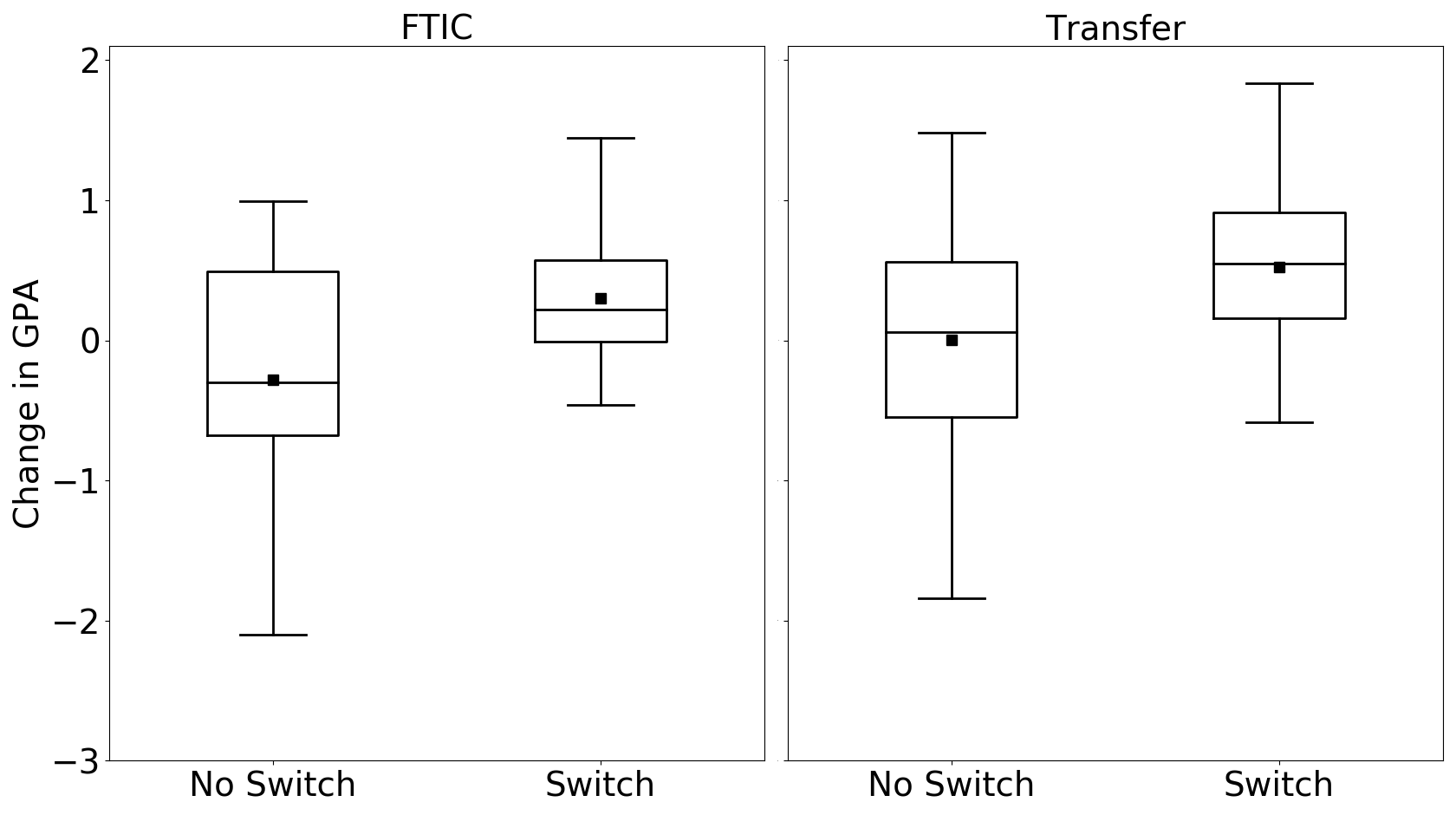}
\caption{comparing changes in GPA between students who switch from MES to FES and students who stay MES for FTIC and transfer students}
\label{fig:FTIC_TR_MES_TO_FES_change_GPA}
\end{figure}

The same kind of analysis is conducted for students who switch their enrollment strategies from MES to FES. Two groups of students are considered: students who change their enrollment strategies after three MES semesters to FES, and students who keep their enrollment strategies as MES. Again, both groups are selected from the same colleges with similar GPAs at the end of the third semester. For those FTIC students who switch their policy in the fourth semester, the average difference in GPA between the fourth semester and mean of the first three semesters is +0.30, while for students who stay MES, this difference is averaged at -0.28. These values for transfer students are  +0.53 and +0.004 for students who do and do not switch, respectively. These results are shown in \figref{fig:FTIC_TR_MES_TO_FES_change_GPA}.

Hypothesis t-test for average GPA before and after switching for both FTIC and transfer students shows that the differences are statistically significant. Similar to the analysis we made for FES to PES strategy switch, we have conducted a two-way ANOVA test to identify if changes in GPA for the student who switches from MES to FES is impacted by their colleges of enrollment. \tabref{tab:ANOVA_FTIC_MES_to_FES} and \tabref{tab:ANOVA_Tr_MES_to_FES} summarize the Anova test results for FTIC and transfer students respectively. Our findings are identical with the previous analysis in that the switching variable (with P-value close to 0) is the only factor affecting students' GPA.

\begin{table}
\centering
\caption{Analysis of variance for the linear regression model for FTIC students switching from MES to FES }
\label{tab:ANOVA_FTIC_MES_to_FES}       
\begin{tabular}{llllll}
\hline\noalign{\smallskip}
Source&DF&Adj SS&Adj MS&F-Value&P-Value\\
\noalign{\smallskip}\hline\noalign{\smallskip}
College &4 &1.442&0.36049&0.43&0.789  \\
Switching &1 &4.9863&4.98631&5.89&0.019 \\
College*Switching &4 &0.2207&0.05518&0.07&0.992  \\
Error &54 &45.6819&0.84596&  &   \\
Total &63 &52.6965 &  & &   \\
\noalign{\smallskip}\hline
\end{tabular}
\end{table}

\begin{table}
\centering
\caption{Analysis of variance for the linear regression model for transfer students switching from MES to FES }
\label{tab:ANOVA_Tr_MES_to_FES}       
\begin{tabular}{llllll}
\hline\noalign{\smallskip}
Source&DF&Adj SS&Adj MS&F-Value&P-Value\\
\noalign{\smallskip}\hline\noalign{\smallskip}
College &5 &0.7707&0.1541&0.24&0.941  \\
Switching &1 &6.1566&6.1566&9.76&0.003 \\
College*Switching &5 &2.8658&0.5732&0.91&0.481  \\
Error &66 &41.6462&0.6310&  &   \\
Total &77 &50.5700 &  & &   \\
\noalign{\smallskip}\hline
\end{tabular}
\end{table}

\textbf{Key Finding 9: All Students:} Switching from FES to PES:

For students who switch from FES to PES, the probability of drop out will be increased, and their GPA will be decreased.

\textbf{Key Finding 10: All Students:} Switching from MES to FES:

For students who switch from FES to PES, the probability of drop out will be decreased, and their GPA will be increased.

\section{Conclusion}
In this paper, we examine the impact of students' enrollment patterns on their academic performance. We applied the Hidden Markov model on our data set collected from the University of Central Florida to classify students based on their enrollment strategies. This classification divides students into three categories: full-time enrollment strategy, part-time enrollment strategy, and mixed enrollment strategy. Assessing academic performance for each enrollment strategy shows that the FES groups have a higher performance compared to the MES and PES group. Also, financial analysis shows a statistically significant difference between family income distributions for students with different enrollment strategies. All these analyses are conducted for FTIC and transfer separately.

The significant contributions of this research are twofold. Firstly, introduce a new definition for full-time and part-time, and mix for students in the long term based on their enrollment strategies. Secondly, our multi-aspect assessments on each group of students emphasize the PES group's vulnerability while encouraging the university to identify such students early during their studies and help them shift towards a mixed enrollment strategy by providing them with financial, educational, and social support.

Some questions remain unanswered: the first one regards students' GPA with the part-time enrollment strategy. In general, FTIC students have a higher GPA than transfer students, while for PES students, we see that transfer students have a better mean GPA than FTIC students. The second question is about changing students' enrollment strategy after some semesters. We saw in this paper that changing enrollment strategy can affect students' GPA and drop up rate. Therefore, it is essential to know why students decide to change their enrollment strategies after some semesters. We will answer these research questions in our future research. 

\bibliographystyle{apacite}

\bibliography{ref}

\section*{Appendices}

\section*{Appendix A}
In the statistic context, Cohen's d is an effect size to show the difference between the means of two samples. The formula for computing Cohen's d is shown in Equation \ref{eq:Cohend}. In the formula, $M_i$ (i=1, 2) and $SD_i$ correspond to the mean and standard deviation of sample $i$. This method, based on the computed $d$ by Equation \ref{eq:Cohend}, assigns qualitative size effect including small, medium, large, very large, and huge for the difference between the means of the two samples. 

\begin{figure}[!h]
\centering
  \includegraphics[width=0.75\textwidth]{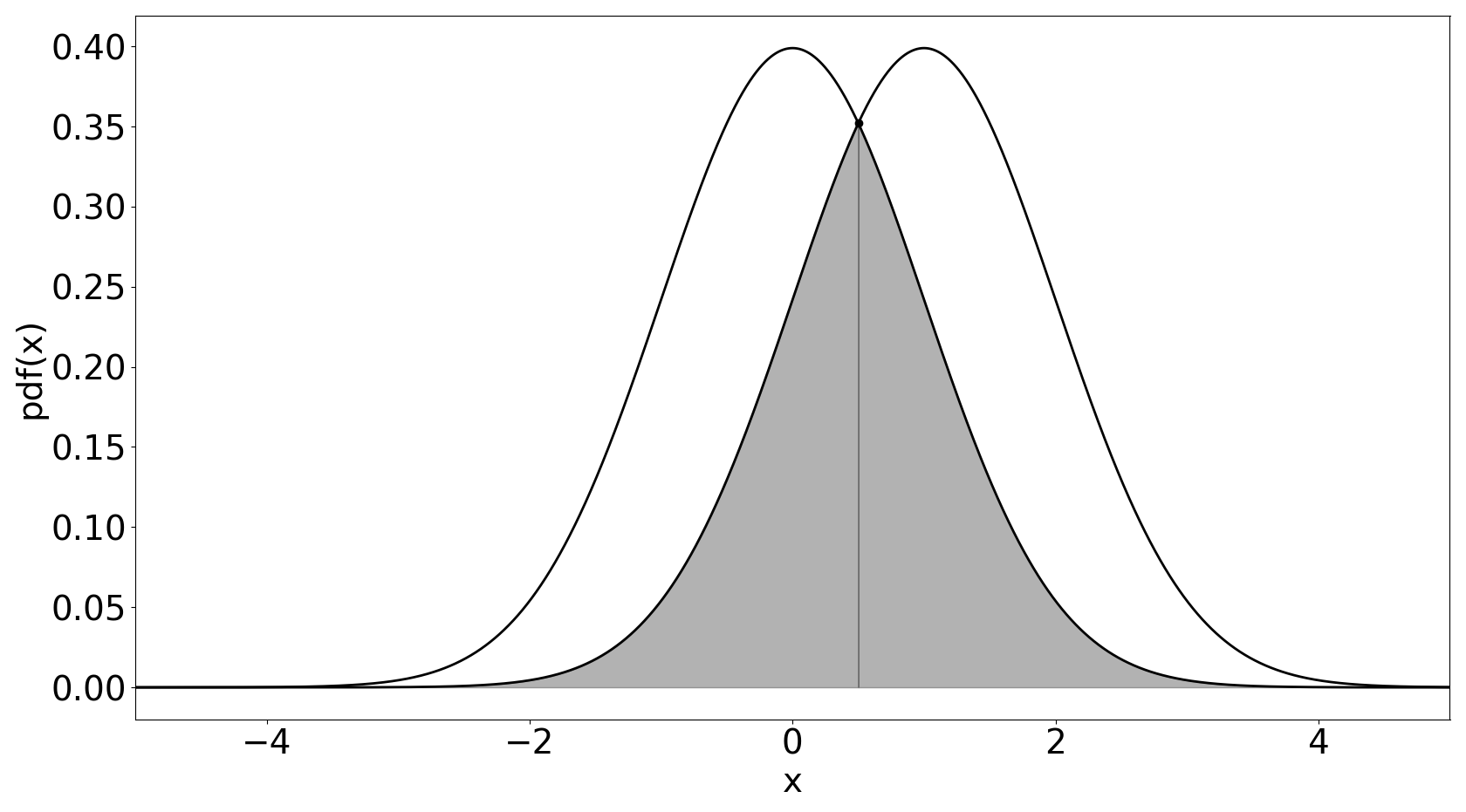}
\caption{Overlap between two normal distributions}
\label{fig:Overlap}       
\end{figure}

\begin{figure}[!h]
\centering
  \includegraphics[width=0.75\textwidth]{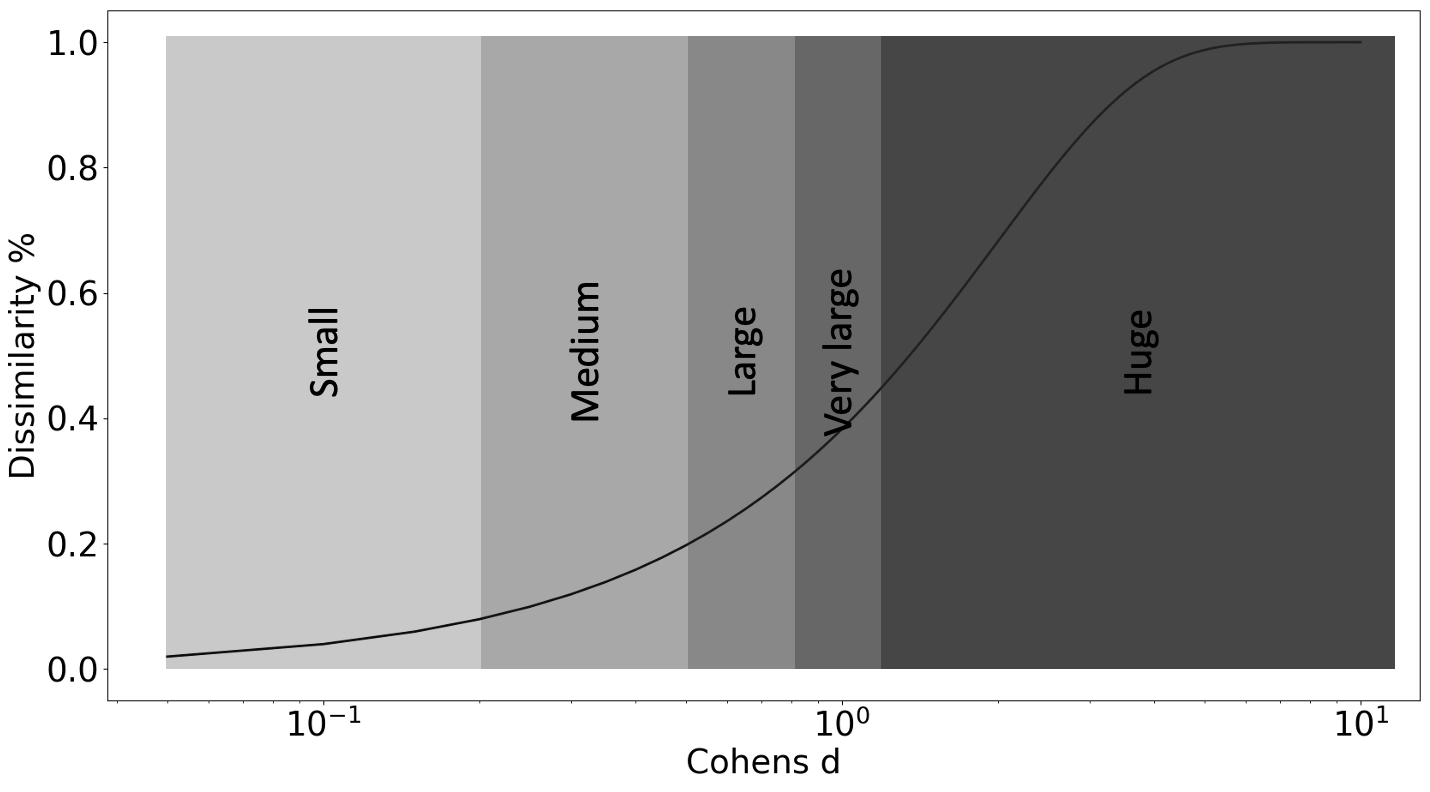}
\caption{Relationship between Cohens d effect size and dissimilarity \%}
\label{fig:dissimilarity}       
\end{figure}

Cohen's d assumes that the two samples have a normal distribution with equal variance. Since students' GPA distribution is negatively skewed, it can affect results accuracy. We use another measurement to quantify the difference between two distribution and name it as \emph{dissimilarity\%}, which is calculated by 1 - \emph{overlap\%} between the two distributions. The \emph{overlap\%} between two normal distributions and relationship between Cohen's d effect size and dissimilarity are illustrated in \figref{fig:Overlap} and \figref{fig:dissimilarity}, respectively. The boundaries for dissimilarity and the corresponding effect size are summarized in \tabref{tab:effect_size_boundry}. 

 \begin{equation} \label{eq:Cohend}
    Cohen's \; d=(M_2-M_1)/{SD_{pooled}}
 \end{equation}

where \begin{equation*} \label{eq:Fundemental_Mat}
    SD_{pooled}=\sqrt{(SD_1^{2}+SD_2^{2})/2}
 \end{equation*}

\begin{table}[!h]
\centering
\caption{Relationship between dissimilarity percentage and effect size}
\label{tab:effect_size_boundry}       
\begin{tabular}{lll}
\hline\noalign{\smallskip}
Dissimilarity \% (D)& Cohen's d&Effect size\\
\noalign{\smallskip}\hline\noalign{\smallskip}
~~~~~~~~~~D $\leq$ 9&d $\leq$ 0.2& Small\\
~~~~~9$ <$ D $\leq$20&0.2$ <$ d $\leq$0.5& Medium \\
~~~~20$ <$ D $\leq$31&0.5$ <$ d $\leq$0.8& Large\\
~~~~31$ <$ D $\leq$42&0.8$ <$ d $\leq$1.2& Very large\\
~~~~42$ <$ D &1.2$ <$ d& Huge\\
\noalign{\smallskip}\hline
\end{tabular}
\end{table}

\section*{Appendix B}
Table \ref{tab:DFW_rate Example} shows the way by which the DFW rate is calculated. DFW rate for each student is defined as the number of courses with D, F, and W grades over the total number of courses taken by the student. After computing the DFW rate for each student, we take average over students' DFW rates in each strategy cluster to have the average DFW rate for students with different enrollment strategies. For example, the average DFW rate for students 1, 2, and 3 is: (0+0.2+.27)/3.

An alternative approach for computing the average DFW rate is that for each strategy cluster, we compute the number of courses with D, F, and W grades over the total number of courses taken by students in each group. Based on this approach, the average DFW rate for students in Table \ref{tab:DFW_rate Example} is: (0+4+8)/(10+20+30)

\begin{table}[htbp]
\caption{Some examples for calculating DFW rate}
\label{tab:DFW_rate Example}       
\begin{tabular}{llll}
\hline\noalign{\smallskip}
Student&\#Num. of courses with DFW&\#Num. of all courses & DFW rate\\
\noalign{\smallskip}\hline\noalign{\smallskip}
~~~~1&~~~~~~~~~~~~~~~~~~~0 & ~~~~~~~~~~~~10& 0$/$10 = 0\\
~~~~2&~~~~~~~~~~~~~~~~~~~4 & ~~~~~~~~~~~~20&4$/$20 = 0.2 \\
~~~~3&~~~~~~~~~~~~~~~~~~~8 & ~~~~~~~~~~~~30&8$/$30 = 0.27\\
\noalign{\smallskip}\hline
\end{tabular}
\end{table}

\end{document}